\documentclass[english,12pt]{article}
\usepackage[T1]{fontenc}
\usepackage{geometry}
\usepackage{verbatim}
\usepackage{float}
\usepackage{bm}
\usepackage{amsmath}
\usepackage{amssymb}
\usepackage{graphicx}

\usepackage{longtable}
\usepackage{subfigure}
\usepackage{mathrsfs}

\usepackage{setspace}

\usepackage{booktabs}
\usepackage{threeparttable}
\usepackage{multirow}

\usepackage[unicode=true,
 bookmarks=false,
 breaklinks=false,pdfborder={0 0 1},colorlinks=false]
 {hyperref}
\hypersetup{
 colorlinks,linkcolor=red,anchorcolor=blue,citecolor=blue}

\makeatletter

\usepackage{cite}\usepackage[]{natbib}\usepackage{amsthm}\usepackage{dsfont}\usepackage{array}\usepackage{mathrsfs}\usepackage{comment}\onecolumn

\usepackage{color}

\newcommand{\bzero}{{\mathbf 0}}

\allowdisplaybreaks

\usepackage{enumitem}
\setlist[itemize]{leftmargin=1.5em}
\setlist[enumerate]{leftmargin=1.5em}

\usepackage[ruled,linesnumbered]{algorithm2e}

\usepackage[pagewise]{lineno}

\newcommand{\roma}{ \bf{\uppercase\expandafter{\romannumeral1}} }
\newcommand{\romb}{ \bf{\uppercase\expandafter{\romannumeral2}} }
\newcommand{\romc}{ \bf{\uppercase\expandafter{\romannumeral3}} }
\newcommand{\romd}{ \bf{\uppercase\expandafter{\romannumeral4}} }
\newcommand{\rome}{ \bf{\uppercase\expandafter{\romannumeral5}} }
\newcommand{\romf}{ \bf{\uppercase\expandafter{\romannumeral6}} }
\newcommand{\romg}{ \bf{\uppercase\expandafter{\romannumeral7}} }

\newcommand{\bu}{\bm{u}}
\newcommand{\bv}{\bm{v}}
\newcommand{\bw}{\bm{w}}

\newcommand{\bA}{\bm{A}}
\newcommand{\bB}{\bm{B}}

\newcommand{\bD}{\bm{D}}

\newcommand{\bI}{\bm{I}}

\newcommand{\bL}{\bm{L}}
\newcommand{\bM}{\bm{M}}

\newcommand{\bT}{\bm{T}}
\newcommand{\bU}{\bm{U}}
\newcommand{\bV}{\bm{V}}

\newcommand{\bX}{\bm{X}}

\newcommand{\0}{\bm{0}}

\newcommand{\cA}{\mathcal{A}}

\newcommand{\cP}{\mathcal{P}}

\newcommand{\cS}{{\mathcal{S}}}

\newcommand{\balpha}{\bm{\alpha}}
\newcommand{\bbeta}{\bm{\beta}}
\newcommand{\bgamma}{\bm{\gamma}}

\newcommand{\bmu}{\bm{\mu}}

\newcommand{\bTheta}{\bm{\Theta}}
\newcommand{\bLambda}{\bm{\Lambda}}

\newcommand{\bSigma}{\bm{\Sigma}}

\def\bbR{\mathbb{R}}

\newcommand{\argmax}{\mathop{\mathrm{argmax}}}

\newcommand{\sign}{\mathop{\mathrm{sign}}}
\newcommand{\tr}{\mathop{\mathrm{tr}}}

\newcommand*{\zero}{{\bm 0}}

\newcommand{\diag}{{\rm diag}}

\newcommand{\ie}{\mbox{\sl i.e.\;}}

\numberwithin{equation}{section}

\definecolor{yxc}{RGB}{255,0,0}
\definecolor{yjc}{RGB}{125,0,0}
\definecolor{cm}{RGB}{0,0,200}
\definecolor{yly}{RGB}{0,150,0}

\newcommand\independent{\protect\mathpalette{\protect\independenT}{\perp}}
\def\independenT#1#2{\mathrel{\rlap{$#1#2$}\mkern2mu{#1#2}}}

\makeatother

\begin{document}

\theoremstyle{plain} \newtheorem{lemma}{\textbf{Lemma}} \newtheorem{prop}{\textbf{Proposition}}\newtheorem{theorem}{\textbf{Theorem}}\setcounter{theorem}{0}
\newtheorem{corollary}{\textbf{Corollary}} \newtheorem{assumption}{\textbf{Assumption}}
\newtheorem{example}{\textbf{Example}} \newtheorem{definition}{\textbf{Definition}}
\newtheorem{fact}{\textbf{Fact}} \newtheorem{condition}{\textbf{Condition}}\theoremstyle{definition}

\theoremstyle{remark}\newtheorem{remark}{\textbf{Remark}}\newtheorem{claim}{\textbf{Claim}}\newtheorem{conjecture}{\textbf{Conjecture}}

\title{Optimal Sparse Sliced Inverse Regression via Random Projection
}
\author{
Jia Zhang\thanks{School of Statistics, Southwestern University of Finance and Economics, Chengdu 611130, China.}
\and Runxiong Wu\thanks{Co-first author. College of Engineering, University of California, Davis, Davis, CA 95616, USA}
\and Xin Chen\thanks{Corresponding author. Department of Statistics and Data Science, Southern University of Science and Technology, Shenzhen 518055, China; Email: \texttt{chenx8@sustech.edu.cn}.}
}

\date{\today}

\maketitle
\begin{abstract}
We propose a novel sparse sliced inverse regression method based on random projections in a large $p$ small $n$ setting. Embedded in a generalized eigenvalue framework, the proposed approach finally reduces to parallel execution of low-dimensional (generalized) eigenvalue decompositions, which facilitates high computational efficiency. Theoretically, we prove that this method achieves the minimax optimal rate of convergence under suitable assumptions. Furthermore, our algorithm involves a delicate reweighting scheme, which can significantly enhance the identifiability of the active set of covariates. Extensive numerical studies demonstrate high superiority of the proposed algorithm in comparison to competing methods.

\end{abstract}


\newpage

\section{Introduction}
Faced with a large number of covariates in various modern applications, Sufficient Dimension Reduction (SDR) provides a statistical framework to reduce the dimension of the problem without loss of information through seeking low-dimensional linear combinations of the original predictors.
In a regression problem involving a univariate response $Y\in\mathbb{R}$ and a $p$-dimensional predictor vector $\bX=(X_1,\ldots, X_p)^{\top}\in\mathbb{R}^p$, SDR aims to find a dimension reduction subspace of $\mathbb{R}^p$ with a basis $\bB$ such that
 \begin{equation}\label{eqn1.1}
 Y\independent \bX \,|\, \bB^\top\bX\,,
 \end{equation}
where $\independent$ stands for statistical independence.
Dimension reduction subspaces are generally not unique, so the primary interest of SDR lies in the intersection of all the dimension reduction subspaces, which enjoys minimum dimensionality and is itself a dimension reduction subspace under mild conditions \citep{cook1996graphics}.
We call the intersection the central subspace and denote it as $\cS_{Y|\bX}$. Let $\bbeta\in\mathbb{R}^{p\times d}$ be a basis of $\cS_{Y|\bX}$, and then $\bbeta^{\top}\bX$ captures the full regression relationship of $Y$ on $\bX$. Since the dimension of the central subspace $d$ is usually much smaller than $p$, then we can use the low-dimensional $\bbeta^T\bX$ to predict $Y$ without loss of any information.

Many SDR methods have been proposed to estimate the central subspace $\cS_{Y|\bX}$ \footnote{See \cite{yin2011sufficient2} and \cite{ma2013review} for through reviews of SDR methods.}, among which the pioneering Sliced Inverse Regression (SIR) \citep{li1991sliced} enjoys high popularity due to its simplicity, generality and computational efficiency. Like most SDR methods, SIR has been proved to be successful in traditional settings where the dimension of the predictors $p$ is fixed or diverges slowly with the sample size $n$ (\citealp{li1991sliced}, \citealp{hsing1992asymptotic}, \citealp{zhu1995asymptotics}, \citealp{zhu2006sliced}). However, when $p\asymp n$ or $p\gg n$, which is quite common in modern datasets, SIR breaks down in both theoretical and computational aspects (\citealp{hung2019sufficient}, \citealp{lin2018on}, \citealp{lin2019sparse}). Indeed, \cite{lin2018on} showed that the SIR estimator of the central subspace is consistent if and only if $p/n$ goes to zero under mild conditions.

To remedy this situation, and to further facilitate interpretability and model parsimony, a reasonable sparsity condition is imposed to restrict the number $s$ of the active predictors in the regression. \cite{lin2018on} proved that the optimal rate for estimating the central subspace should be $(s\log p)/n$ up to some constants in a setting where $\mathrm{Cov}(\bX)=\bI_p$, and proposed a diagonal thresholding algorithm for the single index model which achieves the optimal rate when $\mathrm{Cov}(\bX)=\bI_p$. In a follow-up work, \cite{lin2019sparse} introduced an efficient Lasso variant of SIR for multiple index models with a general covariance matrix of $\bX$. However, this method was shown to be rate optimal when $p$ is of order $o(n^2c^2)$, where $c$ is the generalised signal-to-noise ratio. To fulfil the theoretical gap as well as the restriction on the covariance matrix, \cite{tan2020sparse} proposed an adaptive estimation scheme for sparse SIR, which was computationally tractable and rate optimal even in the case where $\log(p)=o(n)$.

In this paper, we propose a novel efficient Sparse SIR method via Random Projection (SSIRvRP) for large $p$ small $n$ problems. This approach can attain the minmax optimal rate when $\log(p)=o(n)$ under mild assumptions and enjoys high computational efficiency. Compared with existing algorithms,
our SSIRvRP algorithm exhibits marked simplicity, robustness to poor initialization, and scalability through easy parallelization. Indeed, SSIRvRP can be implemented through the parallel execution of low-dimensional (generalized) eigenvalue decompositions, thereby avoiding directly computing the inverse of the sample covariance matrix or the eigenvectors of the sample conditional covariance matrix and selecting multiple tuning parameters appeared in existing sparse SIR methods. Additionally, it does not even require computing the (conditional) sample covariance matrix itself, since it suffices to extract its principal submatrices, which can be computed from the low-dimensional projected data. Finally, the algorithm can be readily extended to other SDR methods via the generalized eigenvalue formulation (\citealp{li2007sparse}, \citealp{hung2019sufficient}).  Extensive numerical experiments demonstrate the superior performance of SSIRvRP in comparison to competing approaches.

When ${\rm Cov}(\bX)=\bI_p$, the proposed algorithm would reduce to the sparse PCA algorithm proposed by \cite{gataric2020sparse}. However, our algorithm is absolutely not a simple and straightforward extension from eigenvalue decomposition to generalized eigenvalue decomposition, which can be reflected from the following three aspects. Firstly and most importantly,
the theoretical evidences behind the proposed algorithm are drastically different from that of \cite{gataric2020sparse}. Our work fulfils the theoretical gap between the standard and generalized eigenvalue decompositions, at least in terms of random projection based algorithms, mainly via wisely utilizing the Cholesky transformation. Secondly, as a generalized eigenvalue problem, the sparse SIR has some unique features, especially in calculation. See Section \ref{sec:alg} for details. Lastly, we embed a reweighting scheme into the proposed algorithm, which significantly improves the ability to identify the active covariates.

{\bf Related literature}.
Random projection techniques have played a role in designing dimension reduction methods. To name a few, \cite{qi2012invariance} and \cite{anaraki2014memory} proposed computing the leading eigenvector of the sample covariance matrix through an ensemble of low-dimensional random projections of the data. \cite{gataric2020sparse} considered sparse principal component analysis via axis-aligned random projections in lager $p$ small $n$ settings, from where we got inspiration for our approach. \cite{hung2019sufficient} introduced a so-called integrated random-partition SDR method by integrating multiple sketches of the central subspace obtained from random partitions of the covariates. Recently, \cite{liu2023random} proposed a random projection approach to hypothesis tests in high-dimensional single index models.

We note that \cite{tan2018sparse} suggested tackling the sparse generalized eigenvalue problem via a truncated Rayleigh flow method. However, this method can only be employed to estimate the leading generalized eigenvector, which limits its application as an SDR approach. For more discussion on sparse SDR methods, please refer to \cite{li2020selective} for a through review.

{\bf Notation}.
We introduce some notation used throughout the paper. For an integer $n > 0$, let $[n]:=\{1,2,\ldots,n\}$ and $\mathbb{E}_n(\bX):=n^{-1}\sum_{i=1}^n\bX_i$ for a random vector $\bX$ with a sample $\{\bX_i\}_{i=1}^n$. For a vector $\balpha\in\mathbb{R}^p$, denote its $j$-th component by $\alpha^{(j)}$, and for $S\subset[p]$, let $\balpha^{(S)}$ denote a subvector of $\balpha$ with components indexed in $S$. Write $\balpha$'s Euclidean norm by $\|\balpha\|_2$.  For a matrix $\bU\in\mathbb{R}^{p\times d}$, let $U^{(i,j)}$ denote its $(i,j)$-th entry, $\bU^{(i,\cdot)}$ denote its $i$-th row, and $\bU^{(\cdot,j)}$ its $j$-th column. For $S\subseteq [p]$ and $S'\subseteq[d]$, write $\bU^{(S,S')}$ be the $|S|\times|S'|$ submatrix with row indexes in $S$ and column indexes in $S'$, and simplify $\bU^{([p],S')}$ and $\bU^{(S,[d])}$ by $\bU^{(\cdot,S')}$ and $\bU^{(S,\cdot)}$, respectively. Let $\|\bU\|_{\mathrm{F}}$ and $\|\bU\|_{\mathrm{op}}$ denote its Frobenius norm and operator norm, respectively.

For any index set $J \subseteq [p]$, $P_J$ signifies the projection matrix which is a $p\times p$ diagonal matrix with the $j$-th diagonal entry being $\mathbf{1}_{\left\{  j\in J \right\}}$. For a real symmetric matrix pair $(\bA,\bB)$ in $\bbR^{p\times p}\times \bbR^{p\times p}$, let $\lambda_{1}(\bA,\bB) \geq \lambda_{2}(\bA,\bB) \geq \ldots \geq \lambda_{p}(\bA,\bB) $ be generalized eigenvalues in decreasing order, and $\bv_1(\bA,\bB),\ldots,\bv_p(\bA,\bB)$ be the corresponding eigenvectors such that $$\bA \bv_i(\bA,\bB)=\lambda_{i}(\bA,\bB)\bB \bv_i(\bA,\bB)$$ for any $i \in [p]$.

\textbf{Organization of the paper}. The rest of the paper is organized as the follows. In Section \ref{sec2}, we propose a sparse SIR estimator via random projection, whose theoretical properties are investigated in Section \ref{sec3}. A reweighting scheme is added to improve the efficiency of the proposed algorithm in Section \ref{sec4}, and Section \ref{sec5} discusses the selection of the hyperparameters for the improved algorithm. Numerical experiments are conducted in Section \ref{sec6}, and in Section \ref{sec7} we applied the proposed method to analyse a gene expression data. Section \ref{sec8} concludes the paper. All the technical proofs are deferred to the Appendix.
  

\section{Method}\label{sec2}

\subsection{Sparse SIR revisited}

Define the discretized version of $Y$ as $$\tilde{Y}=\sum_{h=1}^H h\cdot \mathbf{I}_{\{Y\in J_h\}}\,,$$ where $\{J_1,J_2,\ldots,J_H\}$ is a measurable partition of the sample space of $Y$. If $H\ge d+1$, we know that $\tilde{Y}$ can be used to identify $\cS_{Y|\bX}$ instead of $Y$ with no loss of information (\citealp{BuraCook2001}, \citealp{CookForzani2009}).
Then the SIR procedure is actually a generalized eigenvalue decomposition problem of the kernel matrix $\bSigma_{\mathbb{E}(\bX|Y)} = \mathrm{Cov}\{\mathbb{E}(\bX|\tilde{Y})\}$ with respect to $\bSigma=\mathrm{Cov}(\bX)$, \ie,
\begin{equation}\label{eq:gep}
\bSigma_{\mathbb{E}(\bX|{Y})}\bbeta_i=\lambda_i\bSigma\bbeta_i \, \mbox{ with } \, \bbeta_i^{\top}\bSigma\bbeta_j=\mathbf{1}_{\left\{i=j\right\}}\,,
\end{equation}
where $i,j=1,\cdots,p$, and $\lambda_1\geq\cdots\geq\lambda_d>0=\lambda_{d+1}=\cdots=\lambda_{p}$. {The first $d$ generalized eigenvectors $\left\{ \bbeta_1,\ldots,\bbeta_d \right\}$ corresponding to the nonzero generalized eigenvalues $\lambda_1\geq\cdots\geq\lambda_d$ form a basis of $\cS_{Y|\bX}$.} Thus, solving the following optimization problem yields a basis $\bbeta=(\bbeta_1,\ldots,\bbeta_d)$                                        of $\cS_{Y|\bX}$:
\begin{equation}\label{eqn1.2}
\bbeta = \underset{ \bB\in\bbR^{p\times d} }{\argmax} \, \tr\left\{ \bB^{\top} \bSigma_{\mathbb{E}(\bX| Y)} \bB \right\} \; \mbox{ s.t. } \bB^{\top}\bSigma \bB=\bI_d\,.
\end{equation}

To interpret the extracted components well, it is often encouraged to perform variable selection for SIR. The goal of variable selection is to seek the smallest subset of the predictors $\bX^{(\cA)}$, with partition $\bX=\{(\bX^{(\cA)})^{\top},(\bX^{(\cA^c)})^{\top}\}^{\top}$, such that
\begin{equation}\label{eqn1.3}
Y\independent \bX \,|\,  \bX^{(\cA)}\,,
\end{equation}
where $\cA \subseteq [p]$ denotes the truly relevant predictor set and $\cA^c$ denotes the irrelevant predictor set (\citealp{LiCookNach2005}, \citealp{BondellLi2009}). Following the partition of $\bX$, one can partition $\bbeta$ accordingly as
\begin{equation}\notag
\bbeta=\left(
\begin{array}{c}
\bbeta^{(\cA,\cdot)} \\
\bbeta^{(\cA^c,\cdot)} \\
\end{array}
 \right), \quad \bbeta^{(\cA,\cdot)} \in \bbR^{|\cA|\times d}, \quad \bbeta^{(\cA^c,\cdot)} \in \bbR^{(p-|\cA|)\times d},
\end{equation}
where $|\cA|$ is the cardinality of $\cA$. {Then $(\ref{eqn1.3})$ implies that $\bbeta^{(\cA^c,\cdot)}=\0$ \citep{BondellLi2009}.} Letting $\mbox{supp}(\bbeta)=\{j\in[p]: \bbeta^{(j,\cdot)} \neq \mathbf{0}^\top\}$ be the row support of $\bbeta$, then $\mbox{supp}(\bbeta)=\cA$ in the above partition. Assuming $|\cA|\leq s$, sparse SIR is further defined based on $(\ref{eqn1.2})$ through seeking $\bbeta$ such that
\begin{equation}\label{eqn1.4}
\begin{split}
\bbeta&=\; \underset{ \bB\in\bbR^{p\times d} }{\argmax} \, \tr\left\{ \bB^{\top} \bSigma_{\mathbb{E}(\bX| Y)} \bB \right\}, \\
&\mbox{s.t.}\; \bB^{\top}\bSigma \bB=\bI_d \mbox{ and } |\mbox{supp}(\bB)| \leq s\,.
\end{split}
\end{equation}

A natural sparse SIR estimator can be obtained by solving
\begin{equation}\label{eqn2.1}
\begin{split}
\check{\bbeta}&=\; \underset{ \bB\in\bbR^{p\times d} }{\argmax} \, \tr\left\{ \bB^{\top} \widehat{\bSigma}_{\mathbb{E}(\bX| Y)} \bB \right\}, \\
&\mbox{s.t.}\; \bB^{\top} \widehat{\bSigma} \bB=\bI_d \mbox{ and } |\mbox{supp}(\bB)| \leq s\,,
\end{split}
\end{equation}
where $\widehat{\bSigma}_{\mathbb{E}(\bX| Y)}$ and $\widehat{\bSigma}$ are the sample covariance matrices of the conditional expectation $\mathbb{E}(\bX|\tilde Y)$ and $\bX$, respectively. \cite{tan2020sparse} proved that this natural estimator is rate optimal under various commonly used loss functions. However, solving the problem $(\ref{eqn2.1})$ directly is computationally infeasible as it would require exhaustive search over all $\bB\in\bbR^{p\times d}$ subject to the sparsity constraint. To remedy this problem, \cite{tan2020sparse} further proposed a refined three-steps estimator based on the work of \cite{chao2017sparse}. In the following, we draw inspiration from \cite{gataric2020sparse} and develop another computationally feasible and much simpler estimator based on random projections that achieves optimal statistical rate.

\subsection{A sparse SIR estimator via random projection}\label{sec:alg}

For $k \in [p]$, let $\cP_k := \left\{ P_S: S \subseteq [p] , |S|=k \right\}$ be the set of $k$-dimensional projections.
Our method is described as follows. For two fixed integers $A,B\in \mathbb{N}$, we independently and uniformly generate $A\times B$ projections $\left\{ P_{a,b}: a\in[A],b \in [B] \right\}$ from $\cP_k$. We can treat these projections as $A$ groups, each with cardinality $B$. For each $a \in [A]$, let
\begin{equation}\notag
	b^{*}(a) := \underset{ b \in [B] }{ \mbox{sargmax} } \sum_{i=1}^{d} \lambda_i( P_{a,b}  \widehat{\bSigma}_{\mathbb{E}(\bX|Y)}P_{a,b},  P_{a,b}  \widehat{\bSigma}P_{a,b} )
\end{equation}
denote the index of the selected projection within the $a$-th group, where sargmax denotes the
smallest element in the lexicographic ordering for those argmax values.

Consider the oracle case where $P_{a,b^*(a)} = P_{\cA}$ with $\cA = \mathrm{supp}(\bbeta)$. Then \eqref{eqn1.4} reduces to seeking $\bbeta$ such that
\begin{equation}
\begin{split}
\bbeta&=\; \underset{ \bB\in\bbR^{p\times d} }{\argmax} \, \tr\left\{ \bB^{\top} P_{\cA}\bSigma_{\mathbb{E}(\bX| Y)} P_{\cA} \bB \right\}, \\
&~~~~~~~~~\mbox{s.t.}\; \bB^{\top}P_{\cA}\bSigma P_{\cA}\bB=\bI_d\,,
\end{split}
\end{equation}
which implies that $\bbeta$ can be formed by the leading generalized eigenvectors of the pair $(P_{\cA}\bSigma_{\mathbb{E}(\bX| Y)} P_{\cA}, P_{\cA}\bSigma P_{\cA})$. Hence, the basic idea is to construct a set $\hat{S}$ such that $\cA\subseteq\hat{S}$ with high probability. This explains why $b^*(a)$ is defined as stated above, to a certain extent.

Next, in order to aggregate the information of all $A$ estimators, we compute an importance score $\hat{w}^{(j)}$ for the $j$-th variable, which is defined as
\begin{equation}\notag
\hat{w}^{(j)} := \frac 1A \sum_{a=1}^{A}\sum_{i=1}^{d} ( \hat{\lambda}_{a,b^*(a);i}-\hat{\lambda}_{a,b^*(a);d+1} ) (\hat{v}^{(j)}_{a,b^*(a);i})^2,
\end{equation}
where $\hat{\lambda}_{a,b^*(a);i}$ is the $i$-th generalized eigenvalue of the matrix pair $( P_{a,b^*(a)}  \widehat{\bSigma}_{\mathbb{E}(\bX| Y)}P_{a,b^*(a)}, $ $ P_{a,b^*(a)}  \widehat{\bSigma}P_{a,b^*(a)} ) $, and $\hat{v}^{(j)}_{a,b^*(a);i}$ is the $j$-th element of the corresponding generalized eigenvector. This means that we take account, not just of the frequency with which each co-ordinate is chosen, but also their corresponding magnitudes in the selected eigenvector, as well as an estimate of the signal strength. The estimation procedure is summarized as the following algorithm and we name it as SSIRvRP (Sparse SIR via Random Projections).

\begin{algorithm}[!h]
		\caption{pseudocode of the SSIRvRP algorithm for central subspace estimation}
	\label{alg1}
	\KwIn{$\widehat{\bSigma}_{\mathbb{E}(\bX| Y)}, \widehat{\bSigma}, A, B\in \mathbb{N}, k, l \in [p], d\in [k]$ }
	Generate $\left\{  P_{a,b}: a \in [A], b\in [B] \right\}$ independently and uniformly from $\cP_k$\\
	\For{$a=1,\cdots,A$}{
		\For{$b=1,\cdots, B$}{
			for $i\in [d+1]$, compute $\hat{\lambda}_{a,b;i}:=\lambda_{i}( P_{a,b}  \widehat{\bSigma}_{\mathbb{E}(\bX| Y)}P_{a,b},  P_{a,b}  \widehat{\bSigma}P_{a,b} ) $ and the corresponding generalized eigenvector $\hat{\bv}_{a,b;i}=\bv_i( P_{a,b}  \widehat{\bSigma}_{\mathbb{E}(\bX| Y)}P_{a,b},  P_{a,b}  \widehat{\bSigma}P_{a,b} )$ with $\hat\lambda_{a,b;k+1}=0$		
		}
		Compute $	b^{*}(a) :=  \mbox{sargmax}_{ b \in [B] }  \sum_{i=1}^{d} \hat{\lambda}_{a,b;i} $
	}
	Compute $\widehat{\bw}=(\hat{w}^{(1)},\cdots,\hat{w}^{(p)})^{\top}$ with
	$$\hat{w}^{(j)} := \frac 1 A \sum_{a=1}^{A}\sum_{i=1}^{d} ( \hat{\lambda}_{a,b^*(a);i}-\hat{\lambda}_{a,b^*(a);d+1} ) (\hat{v}^{(j)}_{a,b^*(a);i})^2,~j\in[p]$$  \\
	Let $\hat{S}\subseteq[p]$ be the index set of the $l$ largest components of $\widehat{\bw}$\\
	\KwOut{$\widehat{\bbeta}=(\hat{\bv}_1,\ldots,\hat{\bv}_d)$, where $\hat{\bv}_1,\ldots,\hat{\bv}_d$ are the top $d$ generalized eigenvectors of $( P_{\hat{S}}  \widehat{\bSigma}_{\mathbb{E}(\bX| Y)}P_{\hat{S}},  P_{\hat{S}}  \widehat{\bSigma}P_{\hat{S}} )$ }
	
\end{algorithm}


In Algorithm \ref{alg1}, $\widehat\bSigma = \mathbb{E}_n[\{\bX-\mathbb{E}_n(\bX)\}\{\bX-\mathbb{E}_n(\bX)\}^{\top}]$ and
\begin{align}\label{eq:xy}
\widehat\bSigma_{\mathbb{E}(\bX|Y)} = \sum_{h=1}^H \hat{p}_h \{\mathbb{E}_h(\bX|\tilde{Y}=h)-\mathbb{E}_n(\bX)\}\{\mathbb{E}_h(\bX|\tilde{Y}=h)-\mathbb{E}_n(\bX)\}^{\top},
\end{align}
where $\hat{p}_h=\mathbb{E}_n[\mathbf{1}\{\tilde{Y}=h\}]$ and $\mathbb{E}_h(\bX|\tilde{Y}=h)=\hat{p}_h^{-1}n^{-1}\sum_{i=1}^n \bX_i \mathbf{1}\{\tilde{Y}_i=h\}$.
The positive integers $A$, $B$, $k$, $l$ and $d$ are hyperparameters of the proposed algorithm, whose choices will be analysed in the following theoretical and numerical studies.

\begin{remark}\label{r2}
Another method to estimate $\widehat\bSigma_{\mathbb{E}(\bX|Y)}$ is to utilize the identity $\mathrm{Cov}\{\mathbb{E}(\bX|Y)\}=\mathrm{Cov}(\bX)-\mathbb{E}\{\mathrm{Cov}(\bX|Y)\}$. Then, we can estimate $\widehat\bSigma_{\mathbb{E}(\bX|Y)} = \widehat\bSigma-\widehat\bT$, where
\begin{align}\label{eq:T}
\widehat{\bT} = \frac{1}{H}\sum_{h=1}^{H}\bigg\{\frac{1}{n_h}\sum_{i\in S_h}(\bX_i-\bar{\bX}_{S_h})(\bX_i-\bar{\bX}_{S_h})^\top\bigg\}\,,
\end{align}
where $S_1,\ldots,S_H$ contains the sample indexes associated with the partitioned $Y$ according to its scale, $n_h$ denotes the sample size of the slice $S_h$, and $\bar{\bX}_{S_h}$ denotes the sample mean of this slice. This estimator works as well as the one given above in our numerical experiments.
\end{remark}

Notice that if $\bSigma=\bI_p$, Algorithm \ref{alg1} reduces to the algorithm proposed by \cite{gataric2020sparse} for sparse principal component analysis. However, our algorithm is absolutely not a simple and straightforward extension from eigenvalue decomposition to generalized eigenvalue decomposition, which can be reflected from the following aspects. Firstly and most importantly,
the theoretical evidences behind the proposed algorithm are drastically different from that of \cite{gataric2020sparse}. It is well known that the generalized eigenvalue problem is, in principle, more difficult than the standard one, in both theoretical and computational sides. Several frequently used techniques in standard eigenvalue decompositions, like spectral decomposition, have no counterparts in generalized eigenvalue decomposition problems, which brings great challenges to the theoretical analysis of Algorithm \ref{thm1}. Our work fulfils the gap between the standard and generalized eigenvalue decompositions, at least in terms of random projection based algorithms. See Sections \ref{sec:3.2} and \ref{sec:pfthm2} for theoretical details. Secondly, as a generalized eigenvalue problem, the sparse SIR has some unique features, especially in calculation. See the following discussion.

Step 4 of Algorithm \ref{alg1} involves solving a generalized eigenvalue problem with a nonnegative definite matrix pair $( P_{a,b}  \widehat{\bSigma}_{\mathbb{E}(\bX| Y)}P_{a,b},  P_{a,b}  \widehat{\bSigma}P_{a,b} )$, which is different from common generalized eigenvalue problems where the second matrix in the pair is positive definite (\citealp{li2007sparse}, \citealp{tan2018sparse}, \citealp{hung2019sufficient}). The nonnegativeness of the second matrix $P_{a,b}  \widehat{\bSigma}P_{a,b} $ would lead to multiple solutions of generalized eigenvectors.
To see this clearly,
let $S$ denote the set of the $k$ row indexes corresponding to the nonzero diagonal elements of $P_{a,b}$. Then without loss of generality, the matrix pair turns out to be $(P_S  \widehat{\bSigma}_{\mathbb{E}(\bX| Y)}P_S,  P_{S}  \widehat{\bSigma}P_{S} )$ with
\begin{align*}
P_S  \widehat{\bSigma}_{\mathbb{E}(\bX| Y)}P_S = \left(\begin{array}{cc}
\widehat{\bSigma}_{\mathbb{E}(\bX| Y)}^{(S,S)} & \bzero \\
\bzero & \bzero
\end{array}\right),~
P_S  \widehat{\bSigma}P_S = \left(\begin{array}{cc}
\widehat{\bSigma}^{(S,S)} & \bzero \\
\bzero & \bzero
\end{array}\right)\,,
\end{align*}
and the generalized eigenvalue problem in Step 4 with respect to this pair reduces to
\begin{align}\label{eq:gen2}
\widehat\bSigma_{\mathbb{E}(\bX|{Y})}^{(S,S)}\hat\bv_i^{(S)}=\hat\lambda_i\widehat\bSigma^{(S,S)}\hat\bv_i^{(S)} \, \mbox{ with } \,  \hat\bv_i^{(S),\top}\widehat\bSigma^{(S,S)}\hat\bv^{(S)}_j=\mathbf{1}_{\left\{i=j\right\}}
\end{align}
for $i,j=1,\ldots,k$, where $\hat\lambda_i = \lambda_i(P_S  \widehat{\bSigma}_{\mathbb{E}(\bX| Y)}P_S,  P_{S}  \widehat{\bSigma}P_{S} )$ and $\hat\bv_i=(\hat\bv_i^{(S),\top}, \hat\bv_i^{(S^c),\top})^\top=\bv_i(P_S  \widehat{\bSigma}_{\mathbb{E}(\bX| Y)}P_S,  P_{S}  \widehat{\bSigma}P_{S})$. Since \eqref{eq:gen2} does not impose any restriction for $\hat\bv_i^{(S^c)}$, then the leading $k$ generalized eigenvectors of the original problem have infinitely many solutions. To remedy this situation, we choose $\hat\bv_i=(\hat\bv_i^{(S),\top}, \bzero^\top)^\top$ for $i\in[k]$ for the generalized eigenvalue problem in Step 4. This point is quite different from the eigenvalue decomposition of $P_{S}  \widehat{\bSigma}P_{S}$, which would naturally yield sparse eigenvectors as shown in \cite{gataric2020sparse}.

It remains to solve the reduced low-dimensional generalized eigenvalue problem \eqref{eq:gen2}, which can be solve quickly by traditional algorithms. Notice that if the submatrix $\widehat\bSigma^{(S,S)}$ is invertible, then \eqref{eq:gen2} would further reduce to a low-dimensional eigenvalue decomposition problem. Indeed, $\widehat\bSigma^{(S,S)}$ is invertible with high probability for a properly chosen $k$. Recall that $\widehat\bSigma^{(S,S)}=\mathbb{E}_n[\{\bX^{(S)}-\mathbb{E}_n(\bX^{(S)})\}\{\bX^{(S)}-\mathbb{E}_n(\bX^{(S)})\}^{\top}]$. Then it is invertible with probability approaching to $1$ provided that $k\ll n$ and $\bSigma$ has full rank, which contributes to the computational efficiency of the proposed algorithm. Consequently, the computationally intractable sparse SIR problem \eqref{eqn2.1} can be efficiently solved by conducting low-dimensional eigenvalue decompositions through our proposed SSIRvRP algorithm. Moreover, since $\{P_{a,b}:a\in[A],b\in[b]\}$ are generated randomly from $\mathcal{P}_k$, the $A\times B$ low-dimensional eigenvalue decompositions can be executed in parallel, which further facilitates faster computation. Finally, the matrix pair $(\widehat\bSigma_{\mathbb{E}(\bX|{Y})}^{(S,S)}, \widehat\bSigma^{(S,S)})$ can either be extracted from the pair $(\widehat\bSigma_{\mathbb{E}(\bX|{Y})}, \widehat\bSigma)$ or be estimated from the projected covariates $\bX^{(S)}$. The latter option would be preferable when $p$ is sufficiently large.

It is worthy noting that, different from the convex relaxation algorithms for sparse SIR or SDR (\citealp{tan2018sparse}, \citealp{lin2019sparse}, \citealp{tan2020sparse}), the proposed algorithm is not iterative and thus robust to poor initialization.

Another notable advantage of our method lies in its scalability to other SDR methods. It is well known that the class of inverse regression based SDR methods, including the sliced average
variance estimation \citep{cook1991discussion}, principal Hessian directions (\citealp{li1992principal}, \citealp{cook1996graphics}), directional regression \citep{li2007directional}, among others, can be formulated as a generalized eigenvalue problem. Hence, the proposed SSIRvRP algorithm, as a solution to sparse generalized eigenvalue problems, can be readily applied to these methods to obtain a sparse estimator of the central subspace.


\section{Theoretical analysis}\label{sec3}

\subsection{Upper bound for isotropic covariance}

We consider
the setting where
$(\bX_i,\tilde{Y}_i)_{i=1}^n$ are i.i.d. such that $\bX_i|(\tilde{Y}_i=h)\sim \mathcal{N}_p(\bmu_h,\bSigma_h)$ for $h\in[H]$, which is also assumed in \cite{CookYin2001}, \cite{Cook2007}, \cite{CookForzani2009} and \cite{tan2020sparse}. To simplify the theoretical analysis, we firstly assume $\text{Cov}(\bX)=\bI_p$.
Then, the generalized eigenvalue problem \eqref{eq:gep} reduces to the following eigenvalue problem:
\begin{equation}\notag
\bSigma_{\mathbb{E}(\bX|Y)}\bbeta_i=\lambda_i\bbeta_i, \, \mbox{ with } \bbeta_i^{\top}\bbeta_j=\mathbf{1}_{\left\{i=j\right\}}\,,
\end{equation}
where $\bSigma_{\mathbb{E}(\bX|Y)}$ is the covariance matrix of the conditional expectation $\mathbb{E}(\bX|\tilde Y)$, $i,j=1,\cdots,p$, and $\lambda_1\geq\cdots\geq\lambda_d>0=\lambda_{d+1}=\cdots=\lambda_{p}$. Let $\bLambda = \diag(\lambda_1, \ldots, \lambda_d)$ and recall that $\bbeta = \left( \bbeta_1,\ldots,\bbeta_d \right)$ collects the first $d$ eigenvectors  corresponding to the nonzero eigenvalues $\lambda_1\geq\cdots\geq\lambda_d$. Therefore, we can decompose $\bSigma_{\mathbb{E}(\bX|Y)} = \bbeta \bLambda \bbeta^{\top}$.

The following technical conditions are needed.
\begin{description}
\item[(A1)] $\kappa\lambda\geq\lambda_{1}\geq\cdots\geq\lambda_{d}\geq\lambda> 0$ for some constant $\kappa>1$.
\item[(A2)] $\bbeta \in \bTheta_{p,d,s}(\mu)$, where
$$
\bTheta_{p,d,s}(\mu):= \bigg\{ \bV \in \bTheta_{p,d}, {\rm supp}(\bV)\le s, \frac{\max_{j:\|\bV^{(j,\cdot)}\|_2\neq 0} \|\bV^{(j,\cdot)}\|_2}{\min_{j:\|\bV^{(j,\cdot)}\|_2\neq 0} \|\bV^{(j,\cdot)}\|_2} \le \mu \bigg\}\,,
$$
and $\bTheta_{p,d}$ denotes the set of real $p\times d$ matrices with orthonormal columns.
\end{description}

Assumption (A1) can be seen as a coverage condition (\citealp{Cook2004}, \citealp{YuDongShao2016}), and similar assumptions can be found in the literature (\citealp{CaiMaWu2013}, \citealp{Gao2015}, \citealp{tan2020sparse}).
Assumption (A2) is an incoherence condition by which we require the parameter of interest $\bbeta$ do not have too many non-zero rows, and the non-zero rows should have comparable Euclidean norms. For any $\bbeta \in \bTheta_{p,d,s}(\mu)$, since $\sum_{j\in\cA}\|\bbeta^{(j,\cdot)}\|_2^2 = \|\bbeta\|_{\rm F}^2 = d$, Assumption (A2) implies
\begin{align}\label{eq:8}
\frac{d}{s\mu^2}\le \|\bbeta^{(j,\cdot)}\|_2^2\le\frac{d\mu^2}{s}\,,~~~\forall\,j\in\mathcal{A}\,.
\end{align}

For $\bU, \bV \in \bTheta_{p,d}$, following \cite{gataric2020sparse}, we use the loss function
\begin{align}\label{eq:loss}
L(\bU,\bV):=\|\sin\{\bD(\bU,\bV)\}\|_{\rm F}
\end{align}
to evaluate the distance between $\bU$ and $\bV$, where the sine function acts elementwise, $\bD(\bU,\bV)$ is the $d\times d$ diagonal matrix whose $j$-th diagonal entry is the $j$-th principal angel between $\bU$ and $\bV$, \ie, $\cos^{-1}(\sigma_j)$, where $\sigma_j$ is the $j$-th singular value of $\bU^\top\bV$.

In the following theoretical analysis, $s$ and $p$ are allowed to depend on the sample size $n$, while {$\lambda$, $\mu$, $d$ and $H$ are treated as fixed constants.}
Recall that $\widehat\bbeta$ is the output of Algorithm \ref{alg1} with inputs $\widehat{\bSigma}_{\mathbb{E}(\bX| Y)}$, $\widehat{\bSigma}$, $A$, $B$, $d$, $k$ and $l$. Theorem \ref{thm1} gives an upper bound of the proposed SSIRvRP estimator.

\begin{theorem}\label{thm1}
Suppose Assumptions {\rm(A1)-(A2)} hold. If $k \ge \max\{d+1, s\}$, $l \ge s$, and
\begin{align}\label{eq:9}
16K\sqrt\frac{k\log p }{n} \le \frac{\lambda_d}{s\mu^2}\,,
\end{align}
then it holds that
\begin{align*}
\mathbb{P}\bigg\{L(\widehat\bbeta,\bbeta) \le 2K \sqrt\frac{dl\log(p)}{n\lambda_d^2}\bigg\} \ge 1 - c p^{-3} - p\exp\bigg(-\frac{A\lambda_d^2}{50p^2\mu^8\lambda_1^2}\bigg)\,,
\end{align*}
where $K, c>0$ are some constants.
\end{theorem}

Theorem \ref{thm1} is a generalization of Theorem 1 of \cite{gataric2020sparse}, where they consider the estimation of the principal subspace under a restricted covariance concentration condition
that was introduced in \cite{Wangetal2016}. We extend their results to the problem of central subspace estimation, and the results of Lemma \ref{lem1} in the proof of Theorem \ref{thm1} mimic the restricted covariance concentration condition.
\cite{tan2020sparse} also considered the problem of the central subspace estimation and they proposed an adaptive estimation scheme for sparse SIR, {which achieves an upper bound $\sqrt{s\log(p)/n}$ under the loss function defined in \eqref{eq:loss}}. If $\ell \asymp s$, our SSIRvRP estimator also achieves the same bound up to some constants. 

\subsection{Upper bound for general covariance}\label{sec:3.2}

We go further to the real generalized eigenvalue problem for sparse SIR specified in \eqref{eq:gep}, where the covariance $\bSigma$ of the covariates $\bX$ is not restricted to special structures.
It is well known that the generalized eigenvalue problem is, in principle, more difficult than the standard one, especially in terms of theoretical analysis. Thus, several high-level assumptions would be imposed to highlight the technical difference between the generalized eigenvalue problem tackled in this section and the standard one for the simplified setting where $\bSigma=\bI_p$ addressed in the above section.
The data
{$(\bX_i,\tilde{Y}_i)_{i=1}^n$ are also assumed to be i.i.d.

To be consistent with the logic behind Algorithm \ref{thm1} and to utilize standard properties of eigenvalue decomposition, for the sake of theoretical analysis, we hope to find a proper transformation to reduce the generalized eigenvalue decomposition \eqref{eq:gep} to a standard one, while retaining the row sparsity structure of the basis $\bbeta$. It turns out that the Cholesky decomposition of $\bSigma$ plays a crucial role. For a positive-definite covariance matrix $\bSigma$, using the Cholesky decomposition, we can decompose $\bSigma$ as
\begin{align*}
\bSigma = \bL\bL^\top\,,
\end{align*}
where $\bL$ is a real lower triangular matrix with positive diagonal entries. Let $\bbeta_i = (\bL^\top)^{-1}\bgamma_i$ for $i \in [p]$. Then, \eqref{eq:gep} implies that
\begin{equation}\label{eq:gep2}
\{\bL^{-1}\bSigma_{\mathbb{E}(\bX|{Y})}(\bL^{-1})^\top\}\bgamma_i=\lambda_i\bgamma_i \, \mbox{ with } \, \bgamma_i^{\top}\bgamma_j=\mathbf{1}_{\left\{i=j\right\}}\,,
\end{equation}
where $i,j\in[p]$, and $\lambda_1\geq\cdots\geq\lambda_d>0=\lambda_{d+1}=\cdots=\lambda_{p}$. Denote $\bM = \bL^{-1}\bSigma_{\mathbb{E}(\bX|{Y})}(\bL^{-1})^\top$ and $\bgamma =(\bgamma_1,\ldots, \bgamma_d)$, and then $\bM$ is a real symmetric matrix to which the spectral decomposition can be readily applied, \ie, $\bM = \bgamma\bLambda\bgamma^\top $ with $\bLambda= \mathrm{diag}(\lambda_1,\ldots,\lambda_d)$. In addition, the eigenvalues of $\bM$ are just the generalized eigenvalues of the matrix pair $(\bSigma_{\mathbb{E}(\bX|{Y})},\bSigma)$, and, more importantly, the leading eigenvectors $\bgamma =(\bgamma_1,\ldots, \bgamma_d)$ of $\bM$ precisely retain the row sparsity structure of the generalized eigenvectors $\bbeta = (\bbeta_1,\ldots, \bbeta_d)$ of $(\bSigma_{\mathbb{E}(\bX|{Y})},\bSigma)$. To see this clearly, recall that $\bbeta^\top = (\bbeta^{(\cA,\cdot  ),\top},\bzero^\top)$  and $\bbeta_i = (\bL^\top)^{-1}\bgamma_i$ where $\cA$ represents the set of the active covariates and $\bL$ is a lower triangular matrix. Then, it holds that\footnote{Cholesky decomposition is dependent on the order in which the
variables appear in the random vector $\bX$, and
it works when the variables have a natural ordering.}
\begin{align*}
\bgamma = \bL^\top\bbeta=
\left(\begin{array}{cc}
\bL^{(\cA,\cA),\top} & \bL^{(\cA^c,\cA),\top} \\
\bzero^\top & \bL^{(\cA^c,\cA^c),\top}
\end{array}\right)
\left(\begin{array}{c}
\bbeta^{(\cA,\cdot)} \\
\bzero
\end{array}\right)
=\left(\begin{array}{c}
\bL^{(\cA,\cA),\top}\bbeta^{(\cA,\cdot)} \\
\bzero
\end{array}\right)\,,
\end{align*}
where
\begin{align*}
\bL=\left(\begin{array}{cc}
\bL^{(\cA,\cA)} & \bzero \\
\bL^{(\cA^c,\cA)} & \bL^{(\cA^c,\cA^c)}
\end{array}\right)\,.
\end{align*}
The transformation from the generalized eigenvalue problem \eqref{eq:gep} to the standard one \eqref{eq:gep2} with the row sparsity structure of the leading eigenvectors unchanged makes tractable the theoretical analysis of Algorithm \ref{alg1} under general covariance.

The following assumptions are required.
\begin{description}
\item[(A1')] $\iota\theta\geq\theta_{1}\geq\cdots\geq\theta_{p}\geq\theta> 0$ for some constant $\iota>1$, where $\theta_i$ is the eigenvalue of $\bSigma$.
\item[(A2')] $\bgamma \in \bTheta_{p,d,s}(\mu)$, where $\bTheta_{p,d,s}(\mu)$ is defined in Assumption (A2).
\item[(A3)] Define $\mathcal{S}_k:=\{S\subset[p]:|S|=k\}$ for any $k\in[p]$. Then, 
\begin{align*}
\mathbb{P}\bigg[\sup_{S\in\mathcal{S}_k}{\rm max}\{\|\widehat\bL^{(S,S)}-\bL^{(S,S)}\|_{\rm op},\|\widehat\bM^{(S,S)}-\bM^{(S,S)}\|_{\rm op}\}\le K\sqrt\frac{k\log p}{n}
\bigg]\ge1-c_1p^{-c_2}\,,
\end{align*}
where $\widehat\bSigma=\widehat\bL\widehat\bL^\top$ is the Cholesky decomposition of $\widehat\bSigma$, $\widehat\bM = \widehat\bL^{-1}\widehat\bSigma_{\mathbb{E}(\bX|{Y})}(\widehat \bL^{-1})^\top$ and $K,c_1,c_2>0$ are constants.
\item[(A4)] For any $j\in[p]$ and any ordering of $\bX$,  there exists some $\tau\in(0,1]$ such that $\tau\le (L^{(j,j)})^{-2}\le\tau^{-1}$.
\end{description}

Assumption (A1') is a regularity condition for $\bSigma$, whose eigenvalues are required to be bounded away from zero and infinity.
Assumption (A2') is an incoherence condition for the transformed basis $\bgamma$ by which we require $\bgamma$ do not have too many non-zero rows, and the non-zero rows should have comparable Euclidean norms. Recall that $\bgamma = \bL^\top\bbeta$ and $\bL$ is a lower triangular matrix. Hence, Assumption (A2') is equivalent to requiring that $\bbeta$ do not have too many non-zero rows and that the rows of $\bL^{(\cA,\cA),\top}\bbeta^{(\cA,\cdot)}$ have comparable Euclidean norms.
For any $\bgamma \in \bTheta_{p,d,s}(\mu)$, since $\sum_{j\in\cA}\|\bgamma^{(j,\cdot)}\|_2^2 = \|\bgamma\|_{\rm F}^2 = d$, Assumption (A2') implies
\begin{align}\label{eq:8.1}
\frac{d}{s\mu^2}\le \|\bgamma^{(j,\cdot)}\|_2^2\le\frac{d\mu^2}{s}\,,~~~\forall\,j\in\mathcal{A}\,.
\end{align}
Assumption (A3) is indeed a high-level condition imposed on the sample estimates of $\bSigma$ and $\bSigma_{\mathbb{E}(\bX|{Y})}$. When $\bSigma$ is a diagonal matrix, Assumption (A3) holds under mild conditions similar to those in \cite{tan2020sparse}; see their Lemma 1 for details.
Assumption (A4) is a technical condition imposed to guarantee that the  active variables enjoy higher weights than the inactive ones in the population level. Since all the diagonal elements of $\bL$ are positive, Assumption (A4) seems quite mild. Moreover, this assumption is implied by Assumption (A1') provided that $\bSigma$ is a diagonal matrix. Especially, for the simplified case where $\bSigma = \bI_p$, the above assumptions naturally hold.

For $\bU, \bV \in \bTheta_{p,d}(\bSigma):=\{\bV\in\mathbb{R}^{p\times d}:\bV^\top\bSigma\bV=\bI_d\}$, we use the popular general loss function
\begin{align}\label{eq:loss.1}
L(\bU,\bV):=\|\bU\bU^\top-\bV\bV^\top\|_{\rm F}
\end{align}
to evaluate the distance between linear subspaces; see  \cite{CaiMaWu2013}, \cite{Gao2015} and \cite{tan2020sparse}. We note that, when $\bSigma=\bI_p$, this loss function reduces to the one defined in \eqref{eq:loss} up to a constant $\sqrt{2}$.

In the following, $s$ and $p$ are allowed to depend on the sample size $n$, while $\theta$, $\lambda$, $\mu$, $\tau$, $d$ and $H$ are treated as fixed constants.
Theorem \ref{thm11} gives an upper bound of the proposed SSIRvRP estimator for general covariance.

\begin{theorem}\label{thm11}
Suppose Assumptions {\rm(A1)-(A4)} hold. If $k \ge \max\{d+1, s\}$, $l \ge s$, $K\sqrt{k\log(p)/n}\le{\rm min}\{\lambda_1/(4d),\sqrt{\theta_1},\theta_p/(6\sqrt{\theta_1})\}$, $K\sqrt{l\log(p)/n}\le{\rm min}\{\lambda_d/(2\sqrt{2}),\sqrt{\theta_1},\theta_p/(6\sqrt{\theta_1})\}$ and
\begin{align}\label{eq:9.1}
C\sqrt\frac{k\log p }{n} \le \frac{\tau\lambda_d}{s\mu^2}
\end{align}
for some sufficiently large constant $C>0$,
then it holds that
\begin{align*}
\mathbb{P}\bigg\{L(\widehat\bbeta,\bbeta) \le C' \sqrt\frac{l\log(p)}{n}\bigg\} \ge 1 - c' p^{-c_2} - p\exp\bigg(-\frac{A\tau^4\lambda_d^2}{50p^2\mu^8\lambda_1^2}\bigg)\,,
\end{align*}
where $C',c'>0$ are some constants.
\end{theorem}

Theorem \ref{thm11} generalizes Theorem \ref{thm1} to the general covariance case, showing an upper bound $\sqrt{s\log(p)/n}$ for $l\asymp s$, which echoes latest research results for central subspace estimation (\citealp{lin2019sparse}, \citealp{tan2020sparse}). Although seemingly similar to Theorem \ref{thm1}, the proof of Theorem \ref{thm11} is quite different from that of Theorem \ref{thm1}, which wisely utilizes the Cholesky decomposition and several nice properties of triangular matrices; see Section \ref{sec:pfthm2} for details.

\subsection{Lower bound}

Theorem \ref{thm1} and Theorem \ref{thm11}, together with the following Theorem \ref{thm2}, indicate that the SSIRvRP estimator is minmax optimal up to some logarithmic factor, over all possible sparse SIR estimators, provided that $l\asymp s$. Theorem \ref{thm2} establishes a minmax lower bound among all possible sparse SIR estimators $\widetilde{\bbeta}$, which is similar to the lower bound established in \cite{tan2020sparse}. Different from their methods, we require the central subspace satisfy an incoherence condition. However, we show that this kind of restriction on the parameter space does not make the estimation any easier from the mimnax perspective. For simplicity, we assume $(\bX_i,\tilde{Y}_i)_{i=1}^n$ are i.i.d. such that $\bX_i|(\tilde{Y}_i=h)\sim \mathcal{N}_p(\bmu_h,\bSigma_h)$ for $h\in[H]$ and $\bSigma = \bI_p$.



\begin{theorem}\label{thm2}
Assume that $4(s-1)\le p-1$, $(s-1)\log\{(p-1)/(s-1)\}\ge6$ and $5n\ge4s(s-1)\log\{(p-1)/(s-1)\}$. Then
\begin{align*}
\inf_{\widetilde{\bbeta}}\sup_{\bbeta\in\bTheta_{p,d,s}(3)}\mathbb{E}_{P_{\bbeta}}\{L(\widetilde{\bbeta},\bbeta)\}\gtrsim \sqrt{\frac{s\log(p/s)}{n}}\,,
\end{align*}
{where the expectation is with respect to $(\bX_i, {Y}_i)\sim_{i.i.d.}P_{\bbeta}$.}
\end{theorem}

The bound in Theorem \ref{thm2} is similar to the one established for the estimation of the principal eigenspace \citep{gataric2020sparse}. We generalized their conclusion to the setting of the estimation of the central subspace. Despite similar conclusions, the technical proofs of Theorem \ref{thm2} are quite different from theirs. Moreover, the lower bound established here actually holds beyond the normality assumption of $\bX|
\tilde{Y}$ and the isotropic assumption of the covariance matrix.

\section{Improved algorithm}\label{sec4}

\begin{algorithm}[!h]
		\caption{pseudocode of the SSIRvRP algorithm with reweighting}
	\label{alg2}
	\KwIn{$\widehat{\bSigma}_{\mathbb{E}(\bX|Y)}, \widehat{\bSigma}, A_1,A_2,B_1,B_2\in \mathbb{N}, k,l,l^{'}\in [p], d\in [k]$ }
	Generate $\left\{  P_{a,b}: a \in [A_1], b\in [B_1] \right\}$ independently and uniformly from $\cP_k$\\
	\For{$a=1,\cdots,A_1$}{
		\For{$b=1,\cdots, B_1$}{
			for $i\in [d+1]$, compute $\hat{\lambda}_{a,b;i}:=\lambda_{i}( P_{a,b}  \widehat{\bSigma}_{\mathbb{E}(\bX|Y)}P_{a,b},  P_{a,b}  \widehat{\bSigma}P_{a,b} ) $ and the corresponding generalized eigenvector $\hat{\bv}_{a,b;i}=\bv_i( P_{a,b}  \widehat{\bSigma}_{\mathbb{E}(\bX|Y)}P_{a,b},  P_{a,b}  \widehat{\bSigma}P_{a,b} )$ with $\hat\lambda_{a,b;k+1}=0$		
		}
		Compute $	b^{*}(a) := \underset{ b \in [B_1] }{ \mbox{sargmax} } \sum_{i=1}^{d} \hat{\lambda}_{a,b;i} $
	}
	Compute $\widehat{\bw}=(\hat{w}^{(1)},\cdots,\hat{w}^{(p)})^{\top}$ with
	$$\hat{w}^{(j)} := \frac 1A_1 \sum_{a=1}^{A_1}\sum_{i=1}^{d} ( \hat{\lambda}_{a,b^*(a);i}-\hat{\lambda}_{a,b^*(a);d+1} ) (\hat{v}^{(j)}_{a,b^*(a);i})^2,~j\in[p]$$  \\
	Let $\hat{S}^{'}$ be the index set of the $l^{'}$ largest components of $\widehat{\bw}$. Recompute the $l^{'}$-dimensional vector of weights $\widehat{\bw}'$
	by repeating Steps $1-8$ with the submatrix pair $(\widehat{\bSigma}_{\mathbb{E}(\bX|Y)}^{(\hat{S}^{'},\hat{S}^{'})}, \widehat{\bSigma}^{(\hat{S}^{'},\hat{S}^{'})})$, the projection parameters $ A_2, B_2$ and the newly defined $\mathcal{P}_k:= \left\{ P_S: S \subseteq [l'] , |S|=k \right\}$  \\
	Denote by $\hat{S}$ the index set of the $l$ largest components of $\widehat{\bw}'$   \\
	\KwOut{$\widehat{\bbeta}=(\hat{\bv}_1,\ldots,\hat{\bv}_d)$, where $\hat{\bv}_1,\ldots,\hat{\bv}_d$ are the top $d$ generalized eigenvectors of $( P_{\hat{S}}  \widehat{\bSigma}_{\mathbb{E}(\bX|Y)}P_{\hat{S}},  P_{\hat{S}}  \widehat{\bSigma}P_{\hat{S}} )$ }
	
\end{algorithm}

In this section, in order to enhance the identification ability of the active set of covariates, we add a reweighting step to Algorithm \ref{thm1} to refine the weights corresponding to the largest values of $\widehat{\bw}$. Specifically, let $\hat{S}^{'}$ be the index set of the $l^{'}$ largest components of $\widehat{\bw}$ produced in Algorithm \ref{alg1}. We suggest recomputing the weights of these variables in $\hat{S}^{'}$ by repeating Steps 1-8 in Algorithm \ref{alg1}
with the submatrix pair $(\widehat{\bSigma}_{\mathbb{E}(\bX|Y)}^{(\hat{S}^{'},\hat{S}^{'})}, \widehat{\bSigma}^{(\hat{S}^{'},\hat{S}^{'})})$.
Then we select $l$ indices corresponding to the largest values of $\widehat{\bw}'$ to form a set $\hat{S}$, and output an estimate $\widehat{\bbeta}$ as the first $d$ generalized eigenvectors of $( P_{\hat{S}} \widehat{\bSigma}_{\mathbb{E}(\bX|Y)}P_{\hat{S}}, P_{\hat{S}}  \widehat{\bSigma}P_{\hat{S}} ) $. Pseudo code for this modified SSIRvRP algorithm is given in Algorithm $\ref{alg2}$. We find that the new algorithm with a reweighting step really works well, as shown in the numerical studies.

By similar techniques of Theorem \ref{thm11}, we can prove that the estimate of Algorithm \ref{alg2} has the same upper bound $\sqrt{l\log(p)/n}$ as that of Theorem \ref{thm11}. However, it seems that the reweighting scheme in Step 9 of Algorithm \ref{alg2} helps identify $\cA$, the row support of $\bbeta$, more accurately in the finite sample. There is no surprise if we consider the formation of $\hat{S}$ as a process of variable selection. From this angle, the design of Algorithm \ref{alg2} mimics a two-round selection, which offers an opportunity to reassess the importance of the variables and retrieve the mistakenly deleted ones, often accompanied by weak signals, in the first round.


\begin{remark}\label{r1}
Notice that the estimate of SSIRvRP naturally satisfies the orthogonal constraint $\widehat{\bbeta}^\top\widehat\bSigma\widehat\bbeta=\bI_d$. However, the refined three-steps estimator \citep{tan2020sparse} is optimized by relaxing original constraints and thus can not meet the orthogonal constraint without a normalization step. The Lasso-SIR \citep{lin2019sparse} also does not satisfy this constraint. This can be seen another advantage of the proposed algorithm.

\end{remark}

\section{Choice of hyperparameters}\label{sec5}


In the SSIRvRP algorithm, there are several hyperparameters to be selected before the implementation of the algorithm. We find that the proposed algorithm behaves quite robustly to a wide ranges of combinations of $(A,B)$, $(A_1, B_1)$, $k$ and $l'$, as shown in the following numerical experiments. For the choice of the dimension of the central subspace $d$, several existing methods can be applied to the sparse SIR setting (\citealp{chen2010coordinate}, \citealp{lin2019sparse}), and thus we treat $d$ as a known constant.

The sparsity level $l$ is a key tuning parameter in the proposed method. To choose the tuning parameter, we minimize the following criterion, which has been used in \cite{chen2010coordinate}:
\begin{equation}\notag
-\log\left\{\tr( \widehat{\bbeta}_{l}^{\top} \widehat{\bSigma}_{\mathbb{E}(\bX|Y)} \widehat{\bbeta}_{l}) \right\}+\delta \cdot \mbox{df}_l\,,
\end{equation}
where $\widehat{\bbeta}_{l}$ denotes the solution for $\bbeta$ given the sparsity level $l$, $\mbox{df}_l$ denotes the effective number of parameters, and $\delta=2/n$ for the AIC-type criterion and $\delta=\log(n)/n$ for the BIC-type criterion. Since the number of nonzero rows of $\widehat{\bbeta}_{l}$ is just $l$, we can estimate $\mbox{df}_l$ by $(l-d)\cdot d$. Thus, we choose the best $l$ by minimizing
\begin{equation}\notag
-\log\left\{\tr( \widehat{\bbeta}_{l}^{\top} \widehat{\bSigma}_{\mathbb{E}(\bX|Y)} \widehat{\bbeta}_{l}) \right\}+\delta \cdot (l-d)\cdot d\,.	
\end{equation}

One advantage of the tuning process of $l$ is that it is conducted only in Step 10 of Algorithm \ref{alg2} and Steps 1-9 are computed only once. Furthermore, since $l$ is an integer and its parameter space is countable and finite, the tuning process enjoys higher computational efficiency compared with those tuned on a continuous parameter space.

\section{Simulation studies} \label{sec6}

In this section, we conduct extensive numerical experiments to compare the proposed method with several competitive methods, show the effect of the reweighting step in Algorithm \ref{alg2}, and present some empirical instruction for choice of the hyperparameters in the proposed algorithm.

\subsection{Comparison with existing methods}


We compare our method (Algorithm \ref{alg2}) with the Refined Three-Steps estimator (RTS, hereafter) in \cite{tan2020sparse} which was shown to be rate optimal when $\log p = o(n)$, the Lasso-SIR in \cite{lin2019sparse} and DT-SIR in \cite{lin2017optimality}. Lasso-SIR was proved to be rate optimal when $p=o(n^2)$, and DT-SIR was rate optimal when the covariance matrix of $\bX$ is identity. We describe our method as three types: the first one works with the true sparsity level (SSIRvRP), the second one with the sparsity level tuned by the BIC criterion (SSIRvRP-BIC), and the last one with the sparsity level tuned by the AIC criterion (SSIRvRP-AIC).

To be fair, we copy the simulation settings in \cite{tan2020sparse} and recall the results therein. Five models are considered, \ie,
\begin{description}
\item[Model \roma:] $Y=\bbeta^{\top}\bX+\sin(\bbeta^{\top}\bX)+\epsilon,$
\item[Model \romb:] $Y=2\arctan(\bbeta^{\top}\bX)+\epsilon,$
\item[Model \romc:] $Y=(\bbeta^{\top}\bX)^3+\epsilon,$
\item[Model \romd:] $Y=\sinh(\bbeta^{\top}\bX)+\epsilon,$
\item[Model \rome:] $Y=\exp(\bbeta_1^{\top}\bX)\cdot\sign(\bbeta_2^{\top}\bX)+0.2\epsilon,$
\end{description}
where $\bX\sim \mathcal{N}_p(\0,\bSigma)$,
$\epsilon\sim \mathcal{N}(0,1)$, and $\bX$ and $\epsilon$ are independent. The covariance matrix $\bSigma$ follows the following four structures:
\begin{description}
\item[(1)] Identity covariance: $\bSigma = \bI_p$,
\item[(2)] Dense covariance: $\bSigma = (\sigma_{ij})_{1\le i,j\le p}$ with $\sigma_{ii} = 1$ and $\sigma_{ij}=0.6$ for $i\neq j$,
\item[(3)] Toplitz covariance: $\bSigma = (\sigma_{ij})_{1\le i,j\le p}$ with $\sigma_{ij}=0.5^{|i-j|}$,
\item[(4)] Sparse inverse covariance: $\bSigma$ is the correlation matrix of $\bSigma_0$, and $\bSigma_0^{-1}=(w_{ij})_{1\le i,j\le p}$ with $w_{ij} = \mathbf{1}_{\{i=j\}}+0.5\times\mathbf{I}_{\{|i-j|=1\}} + 0.4\times\mathbf{I}_{\{|i-j|=2\}}$.
\end{description}
The regression coefficients $\bbeta \in\bbR^{p\times 1}$ and $(\bbeta_1,\bbeta_2) \in\bbR^{p\times 2}$ are set to have $5$ nonzero rows whose indexes are randomly chosen from the set $[p]$. And the entries in
these nonzero rows are random numbers drawn from the uniform distribution on the finite set $\left\{-2,-1,0,1,2 \right\}$.

For the hyperparameters of the proposed algorithm, we set $(A_1,B_1)=(900, 300)$, $(A_2,B_2)=(600, 200)$, $k=20$, $l'=50$ and $d$ as its true value.
Since the true parameter does not satisfy the constraint $\bbeta^\top\bSigma\bbeta=\bI_d$, we can not compare these sparse SIR estimates directly under the loss function given in \eqref{eq:loss} or \eqref{eq:loss.1}. Consequently, we use the correlation loss (\citealp{li1991sliced}, \citealp{tan2020sparse}), defined as
\begin{equation}\notag
L_{\rho}(\widehat{\bbeta},\bbeta)= 1-\frac 1d \tr\left\{ (\widehat{\bbeta}^{\top}\bSigma\widehat{\bbeta})^{-1}(\widehat{\bbeta}^{\top}\bSigma\bbeta) (\bbeta^{\top}\bSigma\bbeta)^{-1}  (\bbeta^{\top}\bSigma\widehat{\bbeta}) \right\}\,,
\end{equation}
to measure the efficiency of the estimation and to compare our method with its competing methods.
Each simulation is repeated 100 times, and we summarize the average correlation loss across all combinations of five models and four covariance structures under various $(n,p)$ configurations in Tables \ref{tab1}-\ref{tab4}. For each model setting, the average loss of the optimal algorithm is highlighted by a bold font, and the average loss of the sub-optimal algorithm is highlighted by an italic font.

\begin{table}[!htbp]
\centering
	\caption{The averages of correlation loss with the identity covariance structure.}
	 \label{tab1}
	\bigskip
	\resizebox{\textwidth}{!}{
		\begin{threeparttable}
\begin{tabular}{ccccc|ccc}
\toprule[1.5pt]
$(n,p)$                      & Model & \multicolumn{1}{c}{DT-SIR} & \multicolumn{1}{c}{Lasso-SIR} & \multicolumn{1}{c}{RTS} & \multicolumn{1}{c}{SSIRvRP} & \multicolumn{1}{c}{SSIRvRP-BIC} & \multicolumn{1}{c}{SSIRvRP-AIC} \\ \midrule[1pt]
\multirow{5}{*}{(100,200)} & I     & 0.939                      & 0.230                         & 0.070                        & {\bf 0.015}                       & {\sl 0.024}                           & 0.027                           \\
                           & II    & 0.926                      & 0.389                         & {\bf 0.046}                        & {\sl 0.052}                       & 0.053                           & 0.122                           \\
                           & III   & 0.955                      & 0.163                         & 0.021                        & {\bf 0.002}                       & 0.009                           & {\bf 0.002}                           \\
                           & IV    & 0.923                      & 0.395                         & 0.085                        & {\bf 0.002}                       & 0.008                           & {\bf 0.002}                           \\
                           & V     & 0.769                      & 0.404                         & {\bf 0.066}                        & {\sl 0.090}                       & 0.120                           & 0.112                           \\ \cline{2-8}
\multirow{5}{*}{(100,400)} & I     & 0.954                      & 0.435                         & 0.033                        & {\bf 0.011}                       & {\sl 0.017}                           & 0.032                           \\
                           & II    & 0.949                      & 0.448                         & {\bf 0.015}                        & { 0.092}                       & {\sl 0.081}                           & 0.162                           \\
                           & III   & 0.863                      & 0.355                         & 0.039                        & {\bf 0.002}                       & 0.007                           & {\sl 0.004}                           \\
                           & IV    & 0.888                      & 0.611                         & 0.120                        & {\bf 0.002}                       & 0.005                           & {\sl 0.003}                           \\
                           & V     & 0.683                      & 0.527                         & {\bf 0.091}                        & {\sl 0.138}                       & 0.150                           & 0.170                           \\ \cline{2-8}
\multirow{5}{*}{(100,600)} & I     & 0.902                      & 0.587                         & 0.109                        & {\sl 0.023}                       & {\bf 0.022}                           & 0.039                           \\
                           & II    & 0.879                      & 0.675                         & 0.175                        & {\sl 0.109}                       & {\bf 0.099}                           & 0.196                           \\
                           & III   & 0.968                      & 0.489                         & 0.021                        & {\sl 0.006}                       & {0.009}                           & {\bf 0.005}                           \\
                           & IV    & 0.875                      & 0.756                         & 0.215                        & {\bf 0.003}                       & 0.009                           & {\bf 0.003}                           \\
                           & V     & 0.701                      & 0.561                         & 0.179                        & {\bf 0.140}                       & {\sl 0.142}                           & 0.179                           \\ \hline
\multirow{5}{*}{(200,600)} & I     & 0.997                      & 0.094                         & 0.012                        & {\bf 0.003}                       & {\bf 0.003}                           & 0.018                           \\
                           & II    & 0.930                      & 0.233                         & 0.039                        & {\bf 0.010}                       & {\sl 0.014}                           & 0.091                           \\
                           & III   & 0.944                      & 0.058                         & 0.008                        & {\bf 0.001}                       & {\bf 0.001}                           & {\bf 0.001}                           \\
                           & IV    & 0.970                      & 0.212                         & 0.020                        & {\bf 0.001}                       & {\bf 0.001}                           & {\bf 0.001}                           \\
                           & V     & 0.851                      & 0.242                         & 0.046                        & {\bf 0.032}                       & {\bf 0.039} & 0.100                           \\ \hline
\multirow{5}{*}{(400,600)} & I     & 0.014                      & 0.014                         & 0.009                        & {\bf 0.002}                       & {\bf 0.002}                           & 0.010                           \\
                           & II    & 0.924                      & 0.027                         & 0.013                        & {\bf 0.004}                       & {\sl 0.005} & 0.047                           \\
                           & III   & 0.971                      & 0.004                         & 0.004                        & {\bf 0.000}                       & {\bf 0.000}                           & {\bf 0.000}                           \\
                           & IV    & 0.989                      & 0.021                         & 0.009                        & {\bf 0.000}                       & {\bf 0.000}                           & 0.001                           \\
                           & V     & 0.874                      & 0.045                         & 0.019                        & {\bf 0.008}                       & {\sl 0.009}                           & 0.051       \\ \bottomrule[1.5pt]
\end{tabular}
\end{threeparttable}}
\end{table}

\begin{table}[!htb]
\centering
	\caption{The averages of correlation loss with the dense covariance structure.}
	 \label{tab2}
	\bigskip
	\resizebox{\textwidth}{!}{
		\begin{threeparttable}
\begin{tabular}{ccccc|ccc}
\toprule[1.5pt]
$(n,p)$                      & Model & \multicolumn{1}{c}{DT-SIR} & \multicolumn{1}{c}{Lasso-SIR} & \multicolumn{1}{c}{RTS} & \multicolumn{1}{c}{SSIRvRP} & \multicolumn{1}{c}{SSIRvRP-BIC} & \multicolumn{1}{c}{SSIRvRP-AIC} \\  \midrule[1pt]
\multirow{5}{*}{(100,200)} & I     & 0.472                      & 0.106                         & 0.143                        & {\bf 0.045}                       & 0.054                           & {\sl 0.053}                           \\
                           & II    & 0.766                      & 0.283                         & 0.160                        & {\bf 0.130}                       & {\sl 0.129}                           & 0.175                           \\
                           & III   & 0.904                      & 0.483                         & 0.074                        & {\bf 0.003}                       & 0.024                           & {\sl 0.005}                           \\
                           & IV    & 0.902                      & 0.697                         & 0.298                        & {\bf 0.004}                       & 0.028                           & {\sl 0.010}                           \\
                           & V     & 0.439                      & 0.338                         & {\bf 0.129}                        & {\sl 0.173}                       & 0.199                           & 0.184                           \\ \cline{2-8}
\multirow{5}{*}{(100,400)} & I     & 0.856                      & 0.509                         & 0.074                        & {\sl 0.057}                       & {\sl 0.057}                           & {\bf 0.054}                           \\
                           & II    & 0.828                      & 0.887                         & {\bf 0.019}                        & 0.135                       & {\sl 0.124}                           & 0.194                           \\
                           & III   & 0.256                      & 0.026                         & 0.012                        & {\bf 0.004}                       & 0.023                           & {\sl 0.008}                           \\
                           & IV    & 0.245                      & 0.104                         & 0.081                        & {\bf 0.006}                       & 0.033                           & {\sl 0.010}                           \\
                           & V     & 0.432                      & 0.410                         & {\bf 0.139}                        & {\sl 0.207}                       & 0.208                           & 0.218                           \\ \cline{2-8}
\multirow{5}{*}{(100,600)} & I     & 0.883                      & 0.600                         & 0.301                        & {\sl 0.062}                       & {\bf 0.059}                           & 0.080                           \\
                           & II    & 0.340                      & {\sl 0.159}                         & {\bf 0.131}                        & 0.170                       & 0.160                           & 0.235                           \\
                           & III   & 0.928                      & 0.926                         & 0.284                        & {\bf 0.008}                       & 0.027                           & {\sl 0.019}                           \\
                           & IV    & 0.849                      & 0.544                         & 0.453                        & {\bf 0.007}                       & 0.028                           & {\sl 0.018}                           \\
                           & V     & 0.612                      & 0.425                         & 0.245                        & {\bf 0.195}                       & {\sl 0.197}                           & 0.215                           \\ \cline{2-7} \hline
\multirow{5}{*}{(200,600)} & I     & 0.854                      & 0.187                         & 0.023                        & {\bf 0.007}                       & {\sl 0.012}                           & 0.024                           \\
                           & II    & 0.507                      & 0.085                         & 0.096                        & {\sl 0.036}                       & {\bf 0.034}                           & 0.088                           \\
                           & III   & 0.637                      & 0.028                         & 0.031                        & {\bf 0.002}                       & 0.007                           & {\sl 0.004}                           \\
                           & IV    & 0.946                      & 0.944                         & 0.045                        & {\bf 0.002}                       & 0.009                           & {\sl 0.005}                           \\
                           & V     & 0.548                      & 0.307                         & 0.086                        & {\sl 0.078}                       & {\bf 0.073}                           & 0.101                           \\ \hline
\multirow{5}{*}{(400,600)} & I     & 0.822                      & 0.050                         & 0.005                        & {\bf 0.002}                       & {\sl 0.003}                           & 0.010                           \\
                           & II    & 0.287                      & 0.027                         & 0.020                        & {\bf 0.010}                       & {\bf 0.010}                           & 0.038                           \\
                           & III   & 0.472                      & 0.006                         & 0.019                        & {\bf 0.001}                       & {\bf  0.001}                           & {\bf 0.001}                           \\
                           & IV    & 0.591                      & 0.035                         & 0.038                        & {\bf 0.001}                       & {\bf 0.001}                           & 0.002                           \\
                           & V     & 0.363                      & 0.126                         & 0.027                        & {\bf 0.024}                       & {\sl 0.026}                           & 0.054         \\ \bottomrule[1.5pt]
\end{tabular}
\end{threeparttable}}
\end{table}

\begin{table}[!htb]
\centering
	\caption{The averages of correlation loss with the Toplitz covariance structure.}
	 \label{tab3}
	\bigskip
	\resizebox{\textwidth}{!}{
		\begin{threeparttable}
\begin{tabular}{ccccc|ccc}
\toprule[1.5pt]
$(n,p)$                      & Model & \multicolumn{1}{c}{DT-SIR} & \multicolumn{1}{c}{Lasso-SIR} & \multicolumn{1}{c}{RTS} & \multicolumn{1}{c}{SSIRvRP} & \multicolumn{1}{c}{SSIRvRP-BIC} & \multicolumn{1}{c}{SSIRvRP-AIC} \\ \midrule[1pt]
\multirow{5}{*}{(100,200)} & I     & 0.945                      & 0.186                         & 0.073                        & {\bf 0.013}                       & {\sl 0.019}                           & 0.022                           \\
                           & II    & 0.943                      & 0.297                         & {\bf 0.046}                        & {\sl 0.056}                       & 0.058                           & 0.099                           \\
                           & III   & 0.951                      & 0.088                         & 0.022                        & {\bf 0.002}                       & 0.007                           & {\bf 0.002}                           \\
                           & IV    & 0.915                      & 0.323                         & 0.092                        & {\bf 0.002}                       & 0.006                           & {\sl 0.002}                           \\
                           & V     & 0.794                      & 0.345                         & {\bf 0.061}                        & {\sl 0.109}                       & 0.119                           & 0.117                           \\ \cline{2-8}
\multirow{5}{*}{(100,400)} & I     & 0.952                      & 0.283                         & 0.033                        & {\bf 0.013}                       & {\sl 0.024}                           & 0.034                           \\
                           & II    & 0.919                      & 0.368                         & {\bf 0.017}                        & 0.089                       & {\sl 0.081}                           & 0.155                           \\
                           & III   & 0.917                      & 0.254                         & 0.042                        & {\sl 0.003}                       & 0.005                           & {\bf 0.002}                           \\
                           & IV    & 0.907                      & 0.477                         & 0.141                        & {\bf 0.002}                       & 0.006                           & {\bf 0.002}                           \\
                           & V     & 0.763                      & 0.496                         & {\bf 0.099}                        & {\sl 0.127}                       & 0.133                           & 0.152                           \\ \cline{2-8}
\multirow{5}{*}{(100,600)} & I     & 0.927                      & 0.474                         & 0.105                        & {\bf 0.022}                       & {\sl 0.026}                           & 0.036                           \\
                           & II    & 0.874                      & 0.581                         & 0.187                        & {\sl 0.114}                       & {\bf 0.103}                           & 0.188                           \\
                           & III   & 0.935                      & 0.340                         & 0.021                        & {\bf 0.005}                       & 0.014                           & {\sl 0.006}                           \\
                           & IV    & 0.868                      & 0.692                         & 0.202                        & {\bf 0.003}                       & 0.010                           & {\bf 0.003}                           \\
                           & V     & 0.765                      & 0.528                         & 0.201                        & {\sl 0.142}                       & {\bf 0.140}                          & 0.178                           \\ \hline
\multirow{5}{*}{(200,600)} & I     & 0.932                      & 0.070                         & 0.010                        & {\bf 0.004}                       & {\bf 0.004}                           & 0.018                           \\
                           & II    & 0.926                      & 0.150                         & 0.034                        & {\bf 0.012}                       & {\sl 0.016}                           & 0.084                           \\
                           & III   & 0.831                      & 0.030                         & 0.008                        & {\bf 0.001}                       & {\bf 0.001}                           & {\bf 0.001}                           \\
                           & IV    & 0.938                      & 0.151                         & 0.017                        & {\bf 0.001}                       & {\bf 0.001}                           & 0.002                           \\
                           & V     & 0.603                      & 0.179                         & 0.040                        & {\bf 0.028}                       & {\sl 0.035}                           & 0.079                           \\ \hline
\multirow{5}{*}{(400,600)} & I     & 0.372                      & 0.014                         & 0.008                        & {\bf 0.002}                       & {\bf 0.002}                           & 0.009                           \\
                           & II    & 0.289                      & 0.021                         & 0.012                        & {\bf 0.004}                       & {\sl 0.005}                           & 0.043                           \\
                           & III   & 0.176                      & 0.004                         & 0.003                        & {\bf 0.000}                       & {\bf 0.000}                           & {\bf 0.000}                           \\
                           & IV    & 0.577                      & 0.021                         & 0.010                        & {\bf 0.001}                       & {\bf 0.001}                           & {\bf 0.001}                           \\
                           & V     & 0.199                      & 0.039                         & 0.018                        & {\bf 0.007}                       & {\sl 0.009}                           & 0.045           \\ \bottomrule[1.5pt]
\end{tabular}
\end{threeparttable}}
\end{table}

\begin{table}[!htb]
\centering
	\caption{The averages of correlation loss with the sparse inverse covariance structure.}
	 \label{tab4}
	\bigskip
	\resizebox{\textwidth}{!}{
		\begin{threeparttable}
\begin{tabular}{ccccc|ccc}
\toprule[1.5pt]
$(n,p) $                     & Model & \multicolumn{1}{c}{DT-SIR} & \multicolumn{1}{c}{Lasso-SIR} & \multicolumn{1}{c}{RTS} & \multicolumn{1}{c}{SSIRvRP} & \multicolumn{1}{c}{SSIRvRP-BIC} & \multicolumn{1}{c}{SSIRvRP-AIC} \\ \midrule[1pt]
\multirow{5}{*}{(100,200)} & I     & 0.876                      & 0.153                         & 0.084                        & {\bf 0.015}                       & {\sl 0.023}                           & {\sl 0.023}                           \\
                           & II    & 0.906                      & 0.265                         & {\bf 0.060}                        & 0.100                       & {\sl 0.086}                           & 0.127                           \\
                           & III   & 0.848                      & 0.056                         & 0.024                        & {\bf 0.002}                       & 0.004                           & {\bf 0.002}                           \\
                           & IV    & 0.895                      & 0.280                         & 0.108                        & {\bf 0.002}                       & 0.010                           & {\bf 0.002}                           \\
                           & V     & 0.557                      & 0.267                         & {\bf 0.068}                        & {\sl 0.129}                       & 0.142                           & {\sl 0.129}                           \\ \cline{2-8}
\multirow{5}{*}{(100,400)} & I     & 0.930                      & 0.203                         & 0.040                        & {\sl 0.029}                       & {\bf 0.026}                           & 0.031                           \\
                           & II    & 0.943                      & 0.238                         & {\bf 0.016}                        & 0.108                       & {\sl 0.088}                           & 0.159                           \\
                           & III   & 0.851                      & 0.133                         & 0.039                        & {\bf 0.003}                       & 0.009                           & {\bf 0.003}                           \\
                           & IV    & 0.873                      & 0.327                         & 0.118                        & {\bf 0.003}                       & 0.004                           & {\bf 0.003}                           \\
                           & V     & 0.685                      & 0.550                         & {\bf 0.087}                        & {\sl 0.151}                       & 0.154                           & 0.170                           \\\cline{2-8}
\multirow{5}{*}{(100,600)} & I     & 0.930                      & 0.388                         & 0.147                        & {\sl 0.036}                       & {\bf 0.032}                           & 0.046                           \\
                           & II    & 0.930                      & 0.431                         & 0.183                        & {\sl 0.119}                       & {\bf 0.105}                           & 0.172                           \\
                           & III   & 0.926                      & 0.192                         & 0.023                        & {\sl 0.003}                       & 0.008                           & {\bf 0.002}                           \\
                           & IV    & 0.893                      & 0.576                         & 0.241                        & {\bf 0.002}                       & {\sl 0.006}                           & {\sl 0.006}                           \\
                           & V     & 0.778                      & 0.433                         & {\bf 0.159}                        & 0.191                       & {\sl 0.184}                           & 0.201                           \\ \hline
\multirow{5}{*}{(200,600)} & I     & 0.452                      & 0.033                         & 0.011                        & {\sl 0.005}                       & {\bf 0.004}                           & 0.017                           \\
                           & II    & 0.704                      & 0.082                         & 0.038                        & {\bf 0.012}                       & {\sl 0.016}                           & 0.069                           \\
                           & III   & 0.255                      & 0.010                         & 0.007                        & {\bf 0.001}                       & {\bf 0.001}                           & {\bf 0.001}                           \\
                           & IV    & 0.730                      & 0.086                         & 0.021                        & {\bf 0.001}                       & {\bf 0.001}                           & {\bf 0.001}                           \\
                           & V     & 0.319                      & 0.146                         & 0.048                        & {\bf 0.031}                       & {\sl 0.038}                           & 0.073                           \\ \hline
\multirow{5}{*}{(400,600)} & I     & 0.328                      & 0.016                         & 0.012                        & {\bf 0.002}                       & {\bf 0.002}                           & 0.006                           \\
                           & II    & 0.228                      & 0.023                         & 0.012                        & {\bf 0.003}                       & {\sl 0.004}                           & 0.034                           \\
                           & III   & 0.334                      & 0.007                         & 0.011                        & {\bf 0.001}                       & {\bf 0.001}                           & {\bf 0.001}                           \\
                           & IV    & 0.248                      & 0.017                         & 0.008                        & {\bf 0.001}                       & {\bf 0.001}                           & {\bf 0.001}                           \\
                           & V     & 0.257                      & 0.051                         & 0.018                        & {\bf 0.008}                       & {\sl  0.010}                           & 0.045  \\ \bottomrule[1.5pt]
\end{tabular}
\end{threeparttable}}
\end{table}

It is clear that the proposed method performs generally better than the other methods under various model settings. Specifically, SSIRvRP, SSIRvRP-BIC and SSIRvRP-AIC perform much better than the other methods in Models I, III and IV, and the average loss is decreased by almost an order of magnitude. However, in Models II and V, RTS sometimes performs the best, when the sample size and dimension are both low, \ie, $(n,p)=(100,200)$ and $(100,400)$. With the increase of the sample size and dimension, our SSIRvRP tends to catch up quickly from behind, and the decrease of average loss is significant. Although RTS was proved to be optimal when $\log(p)=o(n)$, which is the same as our method, its finite sample performance is generally inferior to ours. The reason may be that the implementation of our method does not involve any choice of initial points, while other approaches are gradient-based and thus vulnerable to a possible bad choice of initialization.


Additionally, when comparing SSIRvRP-AIC and SSIRvRP-BIC, it was found that SSIRvRP-BIC performed slightly better in Models II and III. However, when considering all five models, neither method was found to be superior. Overall, they demonstrated similar performance. Therefore, either method can be chosen as a suitable working algorithm.

Finally, the average correlation loss decreases as $n$ increases which can be seen from the cases $(n,p)=(100,600), (200, 600)$, and $(400,600)$), and increases as $p$ increases seen from the cases $(n,p)=(100,200), (100, 400)$, and $(100,600)$). These results are fairly consistent with our theoretical findings.


Table \ref{tab5} reports similar results as Table \ref{tab3}, where we use equation \eqref{eq:T} in Remark \ref{r2} to calculate the sample covariance matrix of $\mathbb{E}(\bX|\tilde{Y})$. Both estimates works similarly to each other, so you can choose whichever one you prefer.

\begin{table}[!htb]	
	\centering
	\caption{The averages of correlation loss with the Toplitz covariance structure using \eqref{eq:T} for estimating $\bSigma_{\mathbb{E}(\bX|Y)}$.}
	 \label{tab5}
	\bigskip
	\resizebox{\textwidth}{!}{
		\begin{threeparttable}		
			\begin{tabular}{cccccccc}
				\toprule[1.5pt]
		$(n,p)$&Model&DT-SIR & Lasso-SIR & RTS & SSIRvRP &  SSIRvRP-BIC & SSIRvRP-AIC\cr
				\midrule[1pt]
  $(100, 200)$& I & 0.945 & 0.186  &  0.073  & 0.011  & 0.017  & 0.022  \cr
				& II & 0.943 & 0.297  &  0.046  & 0.060  & 0.055  & 0.120  \cr
				& III & 0.951 & 0.088  &  0.022  & 0.002  & 0.009  & 0.002  \cr
				& IV & 0.915 & 0.323  &  0.092  & 0.002  & 0.008  & 0.003  \cr
				& V & 0.794 & 0.345  &  0.061  & 0.125  & 0.135  & 0.157  \cr \cline{2-8}
 $(100, 400)$& I  &0.952&  0.283  &  0.033  & 0.016  & 0.018  & 0.036 \cr
	           & II  &0.919&  0.368  &  0.017  & 0.119  & 0.111  & 0.166 \cr
	           & III  &0.917& 0.254  &  0.042  & 0.003  & 0.006  & 0.003 \cr
	           & IV  &0.907& 0.477  &  0.141  & 0.003  & 0.008  & 0.003 \cr
	           & V  &0.763& 0.496  &  0.099  & 0.142  & 0.157  & 0.177 \cr			\cline{2-8}
 $(100, 600)$& I  &0.927& 0.474  &  0.105  & 0.024  & 0.026  & 0.038 \cr
               & II  &0.874& 0.581  &  0.187  & 0.104  & 0.098  & 0.186 \cr
               & III  &0.935& 0.340  &  0.021  & 0.003  & 0.004  & 0.003 \cr
               & IV  &0.868& 0.692  &  0.202  & 0.002  & 0.007  & 0.006 \cr
               & V  &0.765& 0.528  &  0.201  & 0.125  & 0.135  & 0.157  \cr \hline
$(200, 600)$ & I  &0.932& 0.070  &  0.010  & 0.004  & 0.003  & 0.018 \cr
			   & II  &0.926& 0.150  &  0.034  & 0.012  & 0.016  & 0.081 \cr
			   & III  &0.831& 0.030  & 0.008   & 0.001  & 0.001  & 0.002 \cr
               & IV  &0.938& 0.151  & 0.017   & 0.001  & 0.001  & 0.001 \cr
               & V  &0.603& 0.179  &  0.040  & 0.031  & 0.038  & 0.093 \cr \hline
$(400, 600)$& I   &0.372& 0.014  & 0.008   & 0.002  & 0.002  & 0.008 \cr
               & II  &0.289& 0.021  & 0.012   & 0.003  & 0.005  & 0.039 \cr
               & III  &0.176& 0.004  & 0.003   & 0.000  & 0.000  & 0.001 \cr
               & IV  &0.577& 0.021  & 0.010   & 0.000  & 0.000  & 0.000 \cr
               & V  &0.199& 0.039  & 0.019   & 0.009  & 0.012  & 0.047 \cr               		
				\bottomrule[1.5pt]
			\end{tabular}	
	\end{threeparttable}}
\end{table}

\begin{figure}[!htb]
\centering\includegraphics[width=\textwidth,height=0.8\textheight]{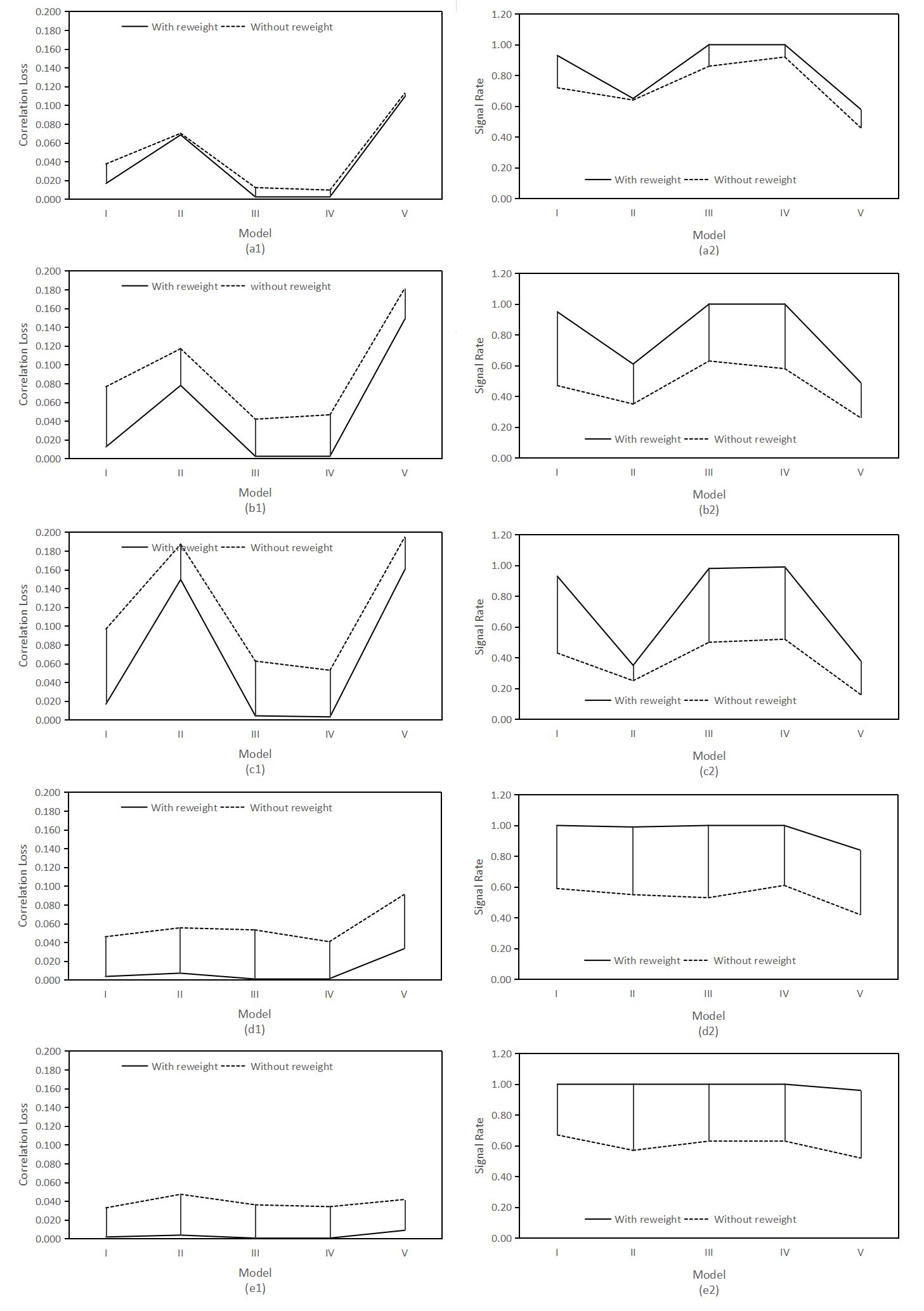}
\caption{Effect of the reweighting step under the Toplitz covariance structure. (a1)-(e1) and (a2)-(e2) correspond to $(n,p)$ settings: $(100,200)$-$(400,600)$.}\label{fig1}
\end{figure}

\begin{figure}[!htb]
\centering\includegraphics[width=\textwidth,height=0.5\textheight]{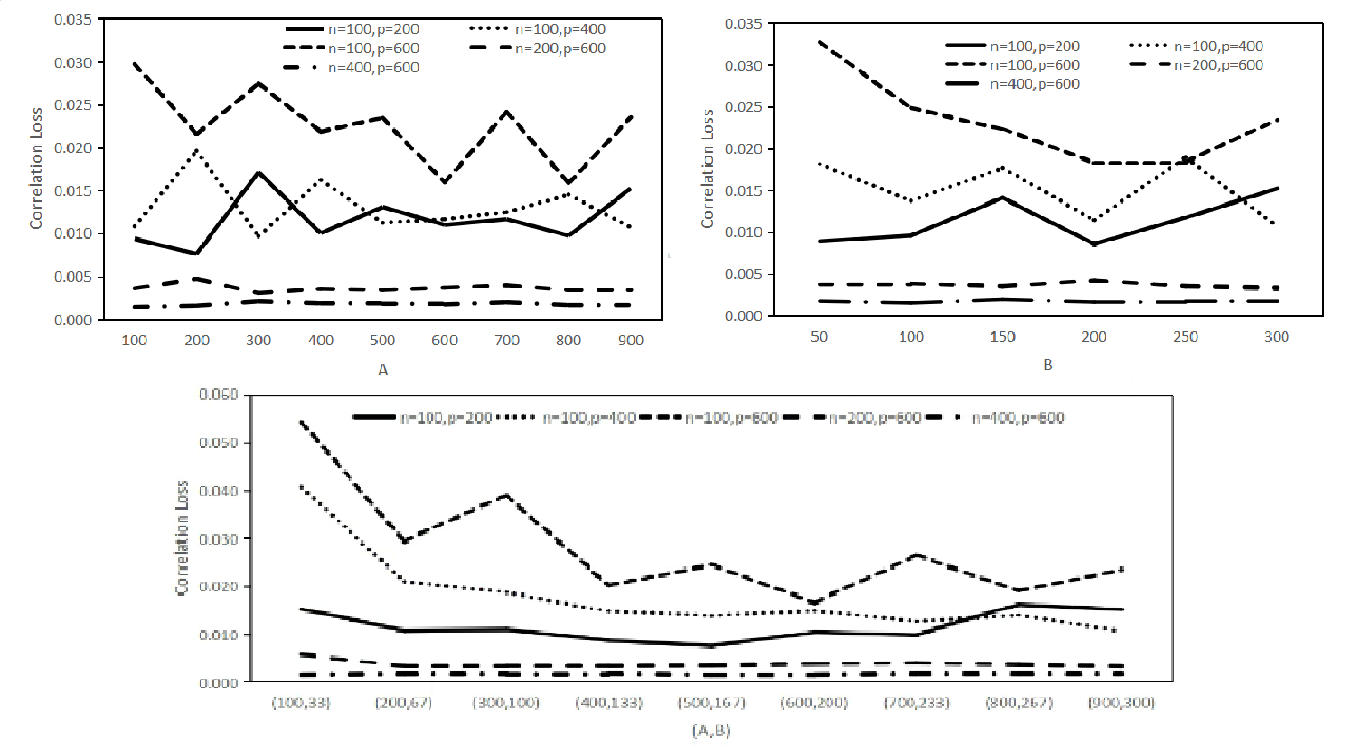}
\caption{Choice of A and B for Model I with Identity covariance.}\label{fig2}
\end{figure}

\begin{figure}[!htb]
\centering\includegraphics[width=0.6\textwidth,height=0.3\textheight]{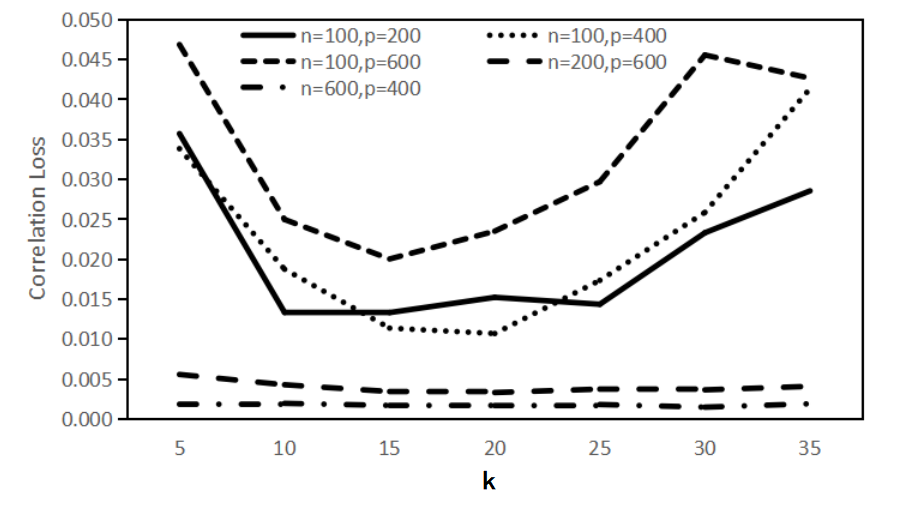}
\caption{Choice of $k$.}\label{fig3}
\end{figure}

\begin{figure}[!htb]
\centering\includegraphics[width=0.6\textwidth,height=0.3\textheight]{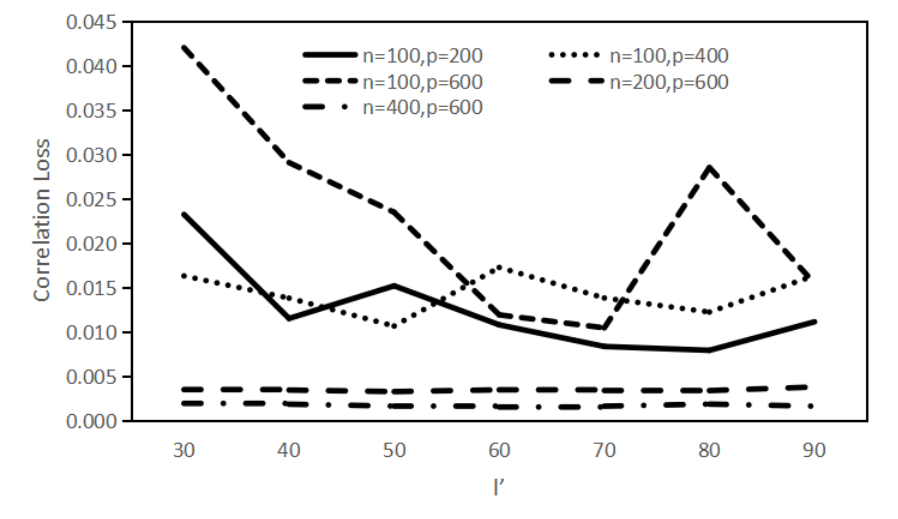}
\caption{Choice of $l'$.}\label{fig4}
\end{figure}

\subsection{The reweighting step}

Compared with Algorithm \ref{alg1}, we add a reweighting step to Algorithm \ref{alg2} to help retrieve true signals. Figure \ref{fig1} displays the effect of the reweighting step in terms of averaged correlation loss and signal rate. Here, signal rate summarizes the  frequency that the algorithm identifies all the true signals exactly. It is clear that the reweighting step significantly decreases the correlation loss and increases the signal rate under various model settings. Therefore, we suggest Algorithm \ref{alg2} for solving sparse SIR problems.

\subsection{Choice of hyperparameters}
In this subsection, we conduct numerical experiments to give some instructions to the choice of the hyperparameters in Algorithm \ref{alg2}. Generally speaking, the performance of the proposed method seems quite robust to a wide range of combinations of the hyperparameters.

Figure \ref{fig2} plots the relationship of the correlation loss and $A$, $B$, and combinations of $(A,B)$ with the other parameters staying unchanged. When $B$ is fixed, increasing $A$ would not significantly decrease the correlation loss, and similar trend occurs when $A$ is fixed. When both $A$ and $B$ are increased simultaneously while keeping $B=A/3$, the figure shows an obvious downward trend for small $A$s. As $A$ gets large to $400$, the decreasing trend of correlation loss becomes flat. Moreover, the loss seems robust to all choices of $A$, $B$ and $(A,B)$ when the sample size increases to $400$ and $600$.

For the choice of $k$, Figure \ref{fig3} shows that their may exist some optimal $k$ for small sample size, while there is no significant trend when $n$ gets large. Figure \ref{fig4} indicates that the loss is robust to the choice of $l'$ when $n=200$. For the cases where $n=100$, the figures shows  a downward trend when $l'$ is small, and the trend gets flattened as $l'$ gets large.

To summarize, when a moderately large number of samples are collected, there is no need to worry about the choice of hyperparameters. However, when the sample size is limited, we suggest lager values for $(A,B)$ and $l'$ and a medium scale for $k$.



\section{Real data study}\label{sec7}


In this section, we consider a gene expression data from the international ``HapMap'' project \citep{thorisson2005international}. The data includes 90 samples, 45 Chinese and 45 Japanese, and reports the gene expression levels from 47293 probes. The gene named CHRNA6 is the subject of many nicotine addiction studies \citep{thorgeirsson2010sequence}, and we treat its mRNA expression as the response $Y$ as done in \cite{fan2018discoveries} and \cite{lin2019sparse}. The expressions of other 47292 genes are treated as the covariates. Consequently, we are now faced with a problem where $p=47292\gg n=90$.

Our goal is to identify several combinations of a few genes that represent the regression relationship of $Y$ regarding $\bX$, and the proposed SSIRvRP method is well-suited for this task. Following \cite{lin2019sparse}, we set $d=1$ and $l=13$ and other hyperparameters as those in the simulation. Based on the estimated coefficients $\widehat\bbeta$, we define $Z=\widehat\bbeta^\top\bX$ and plot $Y$ against $Z$ in Figure \ref{fig5}. 
We then find that there exists a moderate quadratic patten between $Y$ and $Z$. Motivated by this finding, we go further to fit a regression model between $Y$ and $Z,Z^2$, and obtain an adjusted R-squared 0.620 and a p-value 0.000. The mean squared error of the fitted model is 0.040. These results seem a little better than those in \cite{lin2019sparse}, where the authors obtained an R-squared 0.578 and a mean squared error 0.044.  

For comparison, we also employ the sparse PCA algorithm proposed by \cite{gataric2020sparse} to estimate the coefficients $\widetilde\bbeta$ where the information of $Y$ is absent, and calculate $Z' = \widetilde\bbeta^\top\bX$. The scatter plot of $Y$ against $Z'$ does not indicate any interesting patterns between the two variables.
Running a linear regression model between $Y$ and $Z'$ gives us an R-squared 0.001 and a p-value 0.792, indicating that there is no linear patten between the two variables. 
This result seems less meaningful than those achieved by the proposed method which is supervised by $Y$.

\begin{figure}[!htb]
\centering\includegraphics[width=0.6\textwidth,height=0.4\textheight]{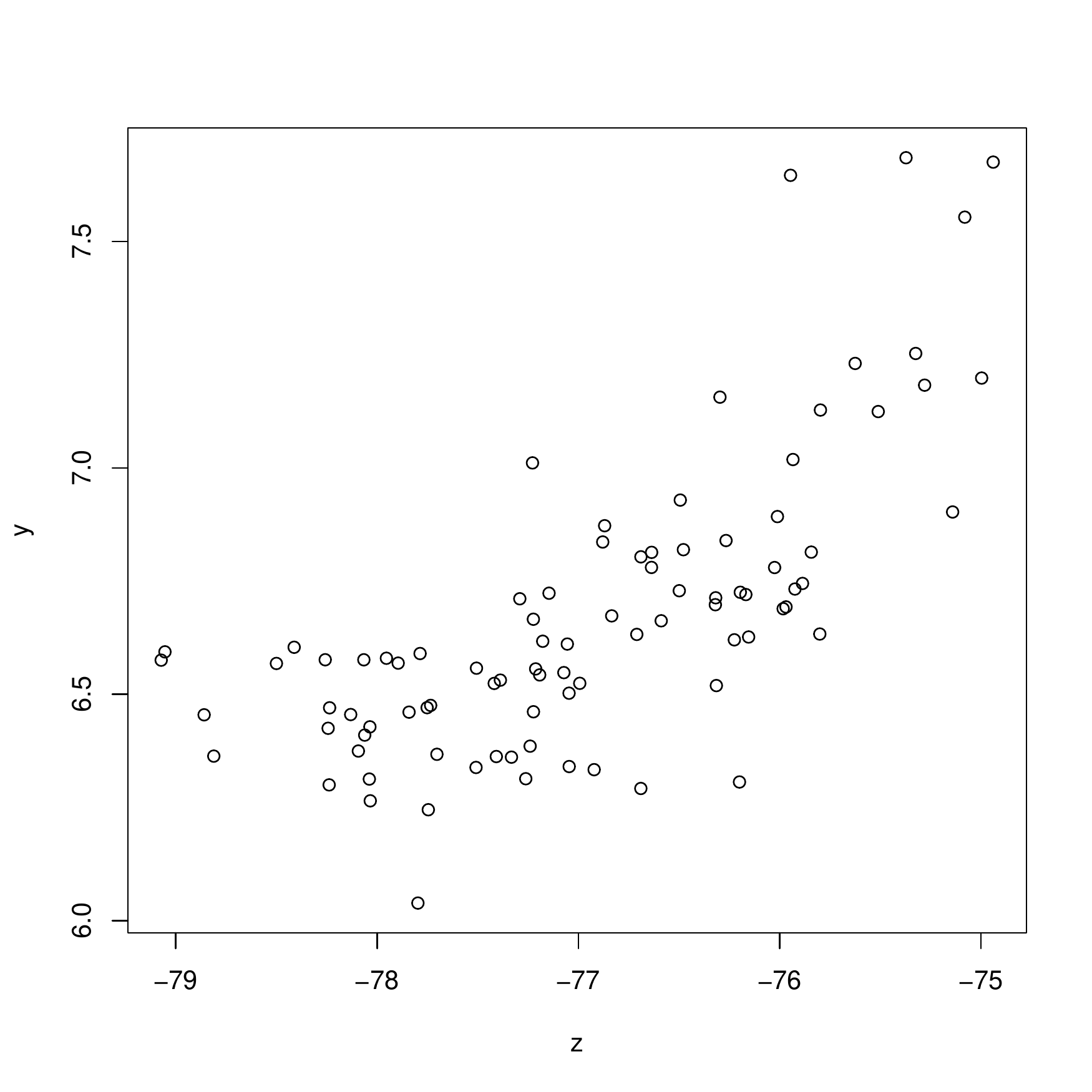}
\caption{Quadratic relationship.}\label{fig5}
\end{figure}

\section{Discussion}\label{sec8}

In this paper, we propose a random projection method to estimate the central subspace in sparse SIR when $p\gg n$. Compared with existing methods, the proposed algorithm is computationally simpler and more efficient. Theoretically, we proved that the proposed estimator achieves the minmax optimal rate under suitable assumptions. It is notable that the random projection technique is introduced to solve a generalize eigenvalue-eigenvector problem. Hence, it can also be applied to sparse Fisher's discriminant analysis for classification and sparse canonical correlation analysis for exploring the relationship of two high-dimensional random vectors. We leave it for further study.


\section{Appendix}

\subsection{Proof of Theorem \ref{thm1}}

The following lemma is needed, whose proof is given in Section \ref{sec:plem1}.

\begin{lemma}\label{lem1}
Under Assumption {\rm (A1)}, there is an event $\Omega$ with probability at least $1-cp^{-3}$ such that on $\Omega$, it holds that
\begin{align*}
{\rm sup}_{\bu\in\mathcal{B}_0^{p-1}(k)} |\bu^{\top}(\widehat{\bSigma}_{\mathbb{E}(\bX|Y)}-\bSigma_{\mathbb{E}(\bX|Y)})\bu| \le K \sqrt\frac{k\log(p)}{n}
\end{align*}
provided that $k \log p \ll n$, where $\mathcal{B}_0^{p-1}(k) = \{\bv=(v^{(1)},\ldots,v^{(p)})^{\top}\in\mathbb{R}^p: \|\bv\|_2 = 1, \sum_{j=1}^p \mathbf{1}\{v^{(j)}\neq0\}\le k\} $ for $k\in[p]$ and $K$ is a constant depending on $H$.
\end{lemma}


We begin the proof.
The proof is similar to that of Theorem 1 of \cite{gataric2020sparse}, but some details should be adjusted, cause we now have a sliced estimator of the covariance matrix.
For notation simplicity, we drop the subscript
$\mathbb{E}(\bX|Y)$ of $\bSigma$, and $\widehat\bSigma$. Define $\mathcal{S}_k:=\{S\subset[p]:|S|=k\}$ and recall $\cA = \{j\in[p]:\bbeta^{(j,\cdot)}\neq\mathbf{0} \}$ which denotes the set of non-zero rows of $\bbeta$. For any $S\in\mathcal{S}_k$, notice that $\bSigma^{(S,S)}=\bbeta^{(S,\cdot)}\bLambda(\bbeta^{(S,\cdot)})^{\top}$. Hence,
\begin{align}\label{eq:12}
\sum_{i=1}^d \lambda_i(\bSigma^{(S,S)})=\sum_{i=1}^k \lambda_i(\bSigma^{(S,S)})-\sum_{i=d+1}^k \lambda_i(\bSigma^{(S,S)}) = \mathrm{tr}(\bSigma^{(S,S)})=\sum_{i=1}^d \sum_{j\in S\cap \mathcal{A}}\lambda_i (\beta^{(j,i)})^2.
\end{align}
Then, by Lemma \ref{lem1}, we know there is an event $\Omega$ with probability at least $1-cp^3$ such that on $\Omega$ we have for any $S\in\mathcal{S}_k$,
\begin{align*}
\|\widehat\bSigma^{(S,S)}-\bSigma^{(S,S)}\|_{\rm op}\le K\sqrt\frac{k\log p}{n}\,,
\end{align*}
where $K>0$ is some constant independent of $S$. On the event $\Omega$ defined above, by Weyl's inequality, we have that
\begin{align*}
\bigg|\sum_{i=1}^d \lambda_i(\widehat\bSigma^{(S,S)}) - \sum_{i=1}^d \sum_{j\in S\cap \mathcal{A}}\lambda_i (\beta^{(j,i)})^2 \bigg| &~\overset{\eqref{eq:12}}{=} \bigg| \sum_{i=1}^d \{ \lambda_i(\widehat\bSigma^{(S,S)}) - \lambda_i(\bSigma^{(S,S)})\}\bigg|\\
&~~~{\le}~Kd\sqrt\frac{k\log p}{n} \overset{\eqref{eq:9}}{\le} \frac{d\lambda_d}{16s\mu^2}\,.
\end{align*}
Thus, on the event $\Omega$, if $(S\cap \mathcal{A}) \subsetneq (S'\cap\mathcal{A})$, then
\begin{align}
&~\sum_{i=1}^d \lambda_i(\widehat\bSigma^{(S,S)})-\sum_{i=1}^d \lambda_i(\widehat\bSigma^{(S',S')}) \notag\\
\le &~ \sum_{i=1}^d \sum_{j\in S\cap \mathcal{A}}\lambda_i (\beta^{(j,i)})^2 - \sum_{i=1}^d \sum_{j\in S'\cap \mathcal{A}}\lambda_i (\beta^{(j,i)})^2 + \frac{d\lambda_d}{8s\mu^2} \notag\\
\le&~ - \sum_{i=1}^d \sum_{j\in (S'\setminus S)\cap \mathcal{A}}\lambda_i (\beta^{(j,i)})^2 + \frac{d\lambda_d}{8s\mu^2}\,. \label{eq:14}
\end{align}
Notice that expression \eqref{eq:8} implies that
\begin{align}
\sum_{i=1}^d \lambda_i (\beta^{(j,i)})^2 \ge \lambda_d \|\bbeta^{(j,\cdot)}\|_2^2 \ge \frac{d\lambda_d}{s\mu^2} \label{eq:14.5}
\end{align}
for any $j\in\mathcal{A}$. Then, combining \eqref{eq:14} and \eqref{eq:14.5}, we have
\begin{align}
\sum_{i=1}^d \lambda_i(\widehat\bSigma^{(S,S)})<\sum_{i=1}^d \lambda_i(\widehat\bSigma^{(S',S')})\,.\label{eq:14.6}
\end{align}

Given the above expression, we now prove that on the event $\Omega$, for some fixed $j\in\mathcal{A}$ and $j'\notin \mathcal{A}$, it holds that
\begin{align}
q_j\ge q_{j'}\,, \label{eq:15}
\end{align}
where $q_k=\mathbb{P}(k\in S_{a,b^*(a)}|\bX)$ for any $k\in[p]$ and some fixed $a\in[A]$. We begin the proof. Define for $\tilde{j}\in\{j,j'\}$ and $b\in[B]$ the following sets:
\begin{align*}
&\mathcal{S}_{b,\tilde{j}}:=\{ (S_{a,1},\ldots,S_{a,B}):b^*(a)=b,\tilde{j}\in S_{a,b} \},\quad\text{and}\\
&~~~~~~~~~\mathcal{S}_{b}:=\{ (S_{a,1},\ldots,S_{a,B}):b^*(a)=b\}\,.
\end{align*}
Let the map $\psi:\mathcal{S}_k\to\mathcal{S}_k$ be defined such that
\begin{align*}
\psi(S):=\bigg\{
\begin{array}{cl}
(S\setminus\{j'\})\cup\{j\} & \text{if}~j'\in S~\text{and}~j\notin S,\\
S & \text{otherwise}.
\end{array}
\end{align*}
Then for every $S\in\mathcal{S}_k$, either $\psi(S)=S$ or $S\cap\mathcal{A}\subsetneq\psi(S)\cap\mathcal{A}$. Hence, inequality \eqref{eq:14.6} implies that
\begin{align}
\sum_{i=1}^d \lambda_i(\widehat\bSigma^{(S,S)})\le\sum_{i=1}^d \lambda_i(\widehat\bSigma^{(\psi(S),\psi(S))})\,.\label{eq:15.1}
\end{align}
Moreover, by the definition of $\psi(\cdot)$, we know that if $j'\in S$ for some $S\in\mathcal{S}_k$, then $j\in\psi(S)$. Thus, for any $(S_{a,1},\ldots,S_{a,b},\ldots,S_{a,B})\in\mathcal{S}_{b,j'}$, there exists $(S_{a,1},\ldots,\psi(S_{a,b}),\ldots,S_{a,B})\in\mathcal{S}_{b,j}$. To see this clearly, since $(S_{a,1},\ldots,S_{a,b},\ldots,S_{a,B})\in\mathcal{S}_{b,j'}$, then $b^*(a)=b$ and $j'\in S_{a,b}$. Notice that $b^*(a)=b$ implies that
\begin{align*}
\max_{b\in[B]}\sum_{i=1}^d \lambda_i(\widehat\bSigma^{(S_{a,b},S_{a,b})})\le\sum_{i=1}^d \lambda_i(\widehat\bSigma^{(S_{a,b},S_{a,b })})\,.
\end{align*}
Together with \eqref{eq:15.1}, we know that
\begin{align*}
\max_{b\in[B]}\sum_{i=1}^d \lambda_i(\widehat\bSigma^{(S_{a,b},S_{a,b})})\le\sum_{i=1}^d \lambda_i(\widehat\bSigma^{(\psi(S_{a,b}),\psi(S_{a,b }))})\,,
\end{align*}
which implies that $(S_{a,1},\ldots,\psi(S_{a,b}),\ldots,S_{a,B})$ satisfies $b^*(a)=b$. Combining the fact that $j\in\psi(S_{a,b})$, we obtain that $(S_{a,1},\ldots,\psi(S_{a,b}),\ldots,S_{a,B})\in\mathcal{S}_{b,j}$. Hence, $|\mathcal{S}_{b,j'}|\le|\mathcal{S}_{b,j}|$. Then, on $\Omega$, it holds for all $b\in[B]$ that
\begin{align*}
&~\mathbb{P}\{j\in S_{a,b^*(a)} |\bX, b^*(a)=b\}\\
=&~\frac{\mathbb{P}\{j\in S_{a,b^*(a)}, b^*(a)=b|\bX\}}{\mathbb{P}\{b^*(a)=b|\bX\}}=\frac{|\mathcal{S}_{b,j}|}{|\mathcal{S}_{b}|}\\
\ge&~\frac{|\mathcal{S}_{b,j'}|}{|\mathcal{S}_{b}|}=\frac{\mathbb{P}\{j'\in S_{a,b^*(a)}, b^*(a)=b|\bX\}}{\mathbb{P}\{b^*(a)=b|\bX\}}\\
=&~\mathbb{P}\{j'\in S_{a,b^*(a)} |\bX, b^*(a)=b\}\,.
\end{align*}
Therefore, we have proved $q_j\ge q_{j'}$ as shown in \eqref{eq:15}.

Notice that
\begin{align*}
\sum_{\tilde{j}\in[p]} q_{\tilde{j}} = \sum_{\tilde{j}\in[p]} \mathbb{P}\{\tilde{j}\in S_{a,b^*(a)}|\bX\}=\frac{1}{B}\sum_{b\in[B]}\sum_{\tilde{j}\in[p]} \mathbb{P}\{\tilde{j}\in S_{a,b^*(a)}|b^*(a)=b,\bX\} = \sum_{\tilde{j}\in[p]} \frac{k}{p} = k\,.
\end{align*}
Thus, by \eqref{eq:15} we obtain on $\Omega$ that
\begin{align}
q_j \ge \frac{ \sum_{\tilde{j}\in([p]\setminus\mathcal{A}\cup\{j\})} q_{\tilde{j}} }{p-s+1}=\frac{k-\sum_{\tilde{j}\in\mathcal{A}\setminus\{j\} } q_{\tilde{j}} }{p-s+1} \ge \frac{k-s+1}{p-s+1} \ge \frac{1}{p}\,.  \label{eq:16}
\end{align}

\begin{remark}
Since $\mathrm{Cov}(\bX)=\bI_p$, $b^{*}(a)$ defined in Algorithm \ref{alg1} reduces to
$$b^{*}(a) :=  \mbox{sargmax}_{ b \in [B] }  \sum_{i=1}^{d} \hat{\lambda}_{a,b;i}\,, $$
where $\hat{\lambda}_{a,b;i} = \lambda_i(P_{a,b}\widehat{\bSigma}P_{a,b})$, the $i$-th largest eigenvalue of the matrix $P_{a,b}\widehat{\bSigma}P_{a,b}$.
\end{remark}

Define $\lambda_{a,b;i}:=\lambda_i(P_{a,b}\bSigma P_{a,b})$ and the corresponding eigenvector $\bv_{a,b;i}:=\bv_i(P_{a,b}\bSigma P_{a,b})$ for $b\in[B]$ and $i\in[p]$.
Notice that $\bSigma$ has $d$ none-zero eigenvalues, so does $P_{a,b} \bSigma P_{a,b}$. Then $\lambda_{a,b;d+1}=\ldots=\lambda_{a,b;p}=0$.
Write $\bV_{a,b} := (\bv_{a,b;1},\ldots,\bv_{a,b:d})$, $\widehat\bV_{a,b} := (\hat\bv_{a,b;1},\ldots,\hat\bv_{a,b:d})$, $\bLambda_{a,b}=\diag(\lambda_{a,b;1},\ldots,\lambda_{a,b;d})$ and $\widehat\bLambda_{a,b}=\diag(\hat\lambda_{a,b;1}-\hat\lambda_{a,b;d+1},\ldots,\hat\lambda_{a,b;d}-\hat\lambda_{a,b;d+1})$. For $\tilde{j} \in \{j,j'\}$, let
\begin{align*}
\hat{w}_a^{(\tilde{j})} := (\widehat\bV_{a,b^*(a)} \widehat\bLambda_{a,b^*(a)} \widehat\bV_{a,b^*(a)}^{\top})^{(\tilde{j},\tilde{j})} = \sum_{i=1}^d (\hat\lambda_{a,b^*(a);i}-\hat\lambda_{a,b^*(a);d+1})(\hat v_{a,b^*(a);i}^{(\tilde{j})})^2.
\end{align*}
We now give upper bound and lower bound of $\hat{w}_a^{(\tilde{j})}$.

Notice that
\begin{align*}
\bV_{a,b^*(a)} \bLambda_{a,b^*(a)} \bV_{a,b^*(a)}^{\top} = \sum_{i=1}^p \lambda_{a,b^*(a);i} \bv_{a,b^*(a);i} \bv_{a,b^*(a);i}^{\top} = P_{a,b^*(a)} \bSigma P_{a,b^*(a)}\,,
\end{align*}
which implies that $(\bV_{a,b^*(a)} \bLambda_{a,b^*(a)} \bV_{a,b^*(a)}^{\top})^{(\tilde{j},\tilde{j})} = \Sigma^{(\tilde{j},\tilde{j})}$ for $\tilde{j}\in S_{a,b^*(a)}$ and $(\bV_{a,b^*(a)} \bLambda_{a,b^*(a)} \bV_{a,b^*(a)}^{\top})^{(\tilde{j},\tilde{j})} = 0 $ otherwise. By Lemma 2 in Appendix A.5 of \cite{gataric2020sparse}, on $\Omega$ we have
\begin{align}
&~| (\widehat\bV_{a,b^*(a)} \widehat\bLambda_{a,b^*(a)} \widehat\bV_{a,b^*(a)}^{\top})^{(\tilde{j},\tilde{j})} - (\bV_{a,b^*(a)} \bLambda_{a,b^*(a)} \bV_{a,b^*(a)}^{\top})^{(\tilde{j},\tilde{j})} | \notag\\
\overset{(1)}{\le} &~ \| \widehat\bV_{a,b^*(a)} \widehat\bLambda_{a,b^*(a)} \widehat\bV_{a,b^*(a)}^{\top} - \bV_{a,b^*(a)} \bLambda_{a,b^*(a)} \bV_{a,b^*(a)}^{\top} \|_{\rm op} \le 4d \| P_{a,b^*(a)}(\widehat\bSigma - \bSigma) P_{a,b^*(a)} \|_{\rm op} \notag\\
\overset{(2)}{\le} &~ 4 K d \sqrt{\frac{k\log(p)}{n}} \overset{\eqref{eq:9}}{\le} \frac{d \lambda_d}{4 s \mu^2}\,, \label{eq:17}
\end{align}
where inequality $(1)$ is obtained by the definition of the
operator norm of a matrix and (2) is implied by Lemma \ref{lem1}. Thus, on $\Omega \cap \{j \in S_{a,b^*(a)}\}$, we have
\begin{align}
\hat{w}_a^{(j)} = &~ (\widehat\bV_{a,b^*(a)} \widehat\bLambda_{a,b^*(a)} \widehat\bV_{a,b^*(a)}^{\top})^{({j},{j})} - (\bV_{a,b^*(a)} \bLambda_{a,b^*(a)} \bV_{a,b^*(a)}^{\top})^{({j},{j})} + (\bV_{a,b^*(a)} \bLambda_{a,b^*(a)} \bV_{a,b^*(a)}^{\top})^{({j},{j})} \notag\\
\ge &~ \Sigma^{(j,j)} - | (\widehat\bV_{a,b^*(a)} \widehat\bLambda_{a,b^*(a)} \widehat\bV_{a,b^*(a)}^{\top})^{({j},{j})} - (\bV_{a,b^*(a)} \bLambda_{a,b^*(a)} \bV_{a,b^*(a)}^{\top})^{({j},{j})} | \overset{\eqref{eq:17}}{\ge} \Sigma^{(j,j)} - \frac{d \lambda_d}{4 s \mu^2} \notag\\
\ge &~ \lambda_d \| \bV^{(j,\cdot)} \|_2^2 - \frac{d \lambda_d}{4 s \mu^2} \overset{\eqref{eq:8}}{\ge} \frac{3 d \lambda_d}{4 s \mu^2}\,. \label{eq:20.1}
\end{align}
Similarly, on $\Omega \cap \{j \in S_{a,b^*(a)}\}$, we have \begin{align}\label{eq:20.2}
\hat{w}_a^{(j)} \le \frac{5 d \lambda_1 \mu^2}{4 s}\,.
\end{align}
Furthermore, on $\Omega \cap \{ j' \in S_{a,b^*(a)} \}$, it holds that
\begin{align}\label{eq:21}
- \frac{d \lambda_d}{4 s \mu^2} \le \hat{w}_a^{(j')} \le \frac{d \lambda_d}{4 s \mu^2}
\end{align}
by noting that $j'\in\mathcal{A}^c$ and $\Sigma^{(j',j')} = \sum_{i=1}^d \lambda_i (\bbeta_i^{(j')})^2 = 0$.
Finally, for all $j\in[p]$, if $j\notin S_{a,b^*(a)}$, then by the definition of $\hat\bv_{a,b^*(a);i}$ for $i\in[d]$, we know that $(\hat v_{a,b^*(a);i})^{(j)}=0$ for $i$ satisfying $\hat\lambda_{a,b^*(a);i}>0$. Thus, by the definition of $\hat{w}_{a}^{(j)}$, it holds that $\hat{w}_{a}^{(j)}=0$ for $j\notin S_{a,b^*(a)}$.
Hence, we have given upper bound and lower bound of $\hat{w}_a^{(\tilde{j})}$.
Using the lower bound and upper bound given above. we obtain on $\Omega$ that
\begin{align}
&~\mathbb{E}(\hat{w}_a^{(j)}-\hat{w}_a^{(j')} | \bX ) \notag\\
= &~ \mathbb{E}[\hat{w}_a^{(j)}(\mathbf{1}\{j\in S_{a,b^*(a)}\} + \mathbf{1}\{j\notin S_{a,b^*(a)}\} ) | \bX ] - \mathbb{E}[\hat{w}_a^{(j')}(\mathbf{1}\{j'\in S_{a,b^*(a)}\} + \mathbf{1}\{j'\notin S_{a,b^*(a)}\} )  | \bX ] \notag\\
=&~ \mathbb{E}(\hat{w}_a^{(j)}\mathbf{1}\{j\in S_{a,b^*(a)}\}-\hat{w}_a^{(j')}\mathbf{1}\{j'\in S_{a,b^*(a)}\} | \bX )\notag \\
\ge &~ \frac{3 q_j d \lambda_d}{4 s \mu^2} -  \frac{q_{j'}d \lambda_d}{4 s \mu^2} \overset{\eqref{eq:15}}{\ge} \frac{ q_j d \lambda_d}{2 s \mu^2} \overset{\eqref{eq:16}}{\ge} \frac{ d \lambda_d}{2p s \mu^2}\,. \label{eq:22}
\end{align}

Now we let $a$, $j$ and $j'$ freely vary again. Define $\Omega_1 := \{\min_{j\in\mathcal{A}} \hat{w}^{(j)} > \max_{j\in\mathcal{A}^c} \hat{w}^{(j)} \}$. Since $l \ge s$, on $\Omega_1$, it holds that $\mathcal{A}\subseteq\hat{S}$, which implies that the $d$ leading eigenvectors of $P_{\hat{S}}\bSigma P_{\hat{S}}$ are the same as those of $\bSigma$.
Hence, by Lemma \ref{lem1} and Theorem 2 of \cite{yu2015useful}, on $\Omega\cap\Omega_1$,
\begin{align*}
L(\widehat\bbeta,\bbeta)\le \frac{2d^{1/2} \|P_{\hat{S}} (\widehat\bSigma-\bSigma) P_{\hat{S}}\|_{\rm op} }{\lambda_d} \le  2K \sqrt\frac{dl\log(p)}{n\lambda_d^2}\,.
\end{align*}
Then it suffices to derivative the lower bound of $\mathbb{P}(\Omega\cap\Omega_1)$. Observe that $\hat{w}^{(j)}=A^{-1}\sum_{a=1}^A \hat{w}_a^{(j)}$ for any $j\in[p]$. Then, for any $j\in\mathcal{A}$ and $j'\in\mathcal{A}^c$, on $\Omega$, it holds that
\begin{align*}
&~~~~~~~~\hat{w}^{(j)}-\hat{w}^{(j')} \\
&~~~~= \{ \hat{w}^{(j)} - \mathbb{E}(\hat{w}^{(j)}|\bX) \} - \{ \hat{w}^{(j')} - \mathbb{E}(\hat{w}^{(j')}|\bX) \} +  \mathbb{E}(\hat{w}^{(j)} - \hat{w}^{(j')} |\bX) \\
&~~~\overset{\eqref{eq:22}}{\ge}  \frac{ d \lambda_d}{2p s \mu^2} + \{ \hat{w}^{(j)} - \mathbb{E}(\hat{w}^{(j)}|\bX) \} - \{ \hat{w}^{(j')} - \mathbb{E}(\hat{w}^{(j')}|\bX) \}\,,
\end{align*}
which implies
\begin{align*}
\Omega_1^c \subseteq \bigg[\bigcup_{j\in\mathcal{A}} \bigg\{ \hat{w}^{(j)} - \mathbb{E}(\hat{w}^{(j)}|\bX) \le \frac{ d \lambda_d}{4p s \mu^2} \bigg\}\bigg] \cup \bigg[\bigcup_{j'\notin\mathcal{A}} \bigg\{ \hat{w}^{(j')} - \mathbb{E}(\hat{w}^{(j')}|\bX) \ge \frac{ d \lambda_d}{4p s \mu^2} \bigg\}\bigg]\,.
\end{align*}
Notice that conditioned on $\bX$, $\{\hat{w}_a^{(j)}\}_{a\in[A]}$ are i.i.d. random variables, and \eqref{eq:20.1}, \eqref{eq:20.2} and \eqref{eq:21} implies that $\hat{w}_a^{(j)}$ is bounded on $\Omega$ for all $j\in[p]$. Thus, by the union bound and the Hoeffding's inequality conditioned on $\bX$, on $\Omega$ we obtain that
\begin{align*}
\mathbb{P}(\Omega_1^c|\bX) \le p\exp\bigg(-\frac{A\lambda_d^2}{50p^2\mu^8\lambda_1^2}\bigg)\,.
\end{align*}
We complete the proof by noting that
\begin{align*}
\mathbb{P}(\Omega\cap\Omega_1) \ge 1 - c p^{-3} - p\exp\bigg(-\frac{A\lambda_d^2}{50p^2\mu^8\lambda_1^2}\bigg)\,.
\end{align*}
\hfill$\Box$


\subsection{Proof of Theorem \ref{thm11}}
\label{sec:pfthm2}
Recall $\mathcal{S}_k:=\{S\subset[p]:|S|=k\}$ and $\cA = \{j\in[p]:\bbeta^{(j,\cdot)}\neq\mathbf{0} \}$. For any $S\in\mathcal{S}_k$, notice that $\bM^{(S,S)}=\bgamma^{(S,\cdot)}\bLambda(\bgamma^{(S,\cdot)})^{\top}$. Hence,
\begin{align}\label{eq:12.1}
&~\sum_{i=1}^d \lambda_i(\bSigma^{(S,S)}_{\mathbb{E}(\bX|Y)},\bSigma^{(S,S)})=\sum_{i=1}^d \lambda_i(\bM^{(S,S)})\\\label{eq:12.2}
=&~\sum_{i=1}^k \lambda_i(\bM^{(S,S)})-\sum_{i=d+1}^k \lambda_i(\bM^{(S,S)}) = \mathrm{tr}(\bM^{(S,S)})=\sum_{i=1}^d \sum_{j\in S\cap \mathcal{A}}\lambda_i (\gamma^{(j,i)})^2\,,
\end{align}
where equation \eqref{eq:12.1} is obtained by the following lemma.

\begin{lemma}\label{lem2}
For any $S\in\mathcal{S}_k$, let $\bSigma^{(S,S)}=\bL_{(S,S)}\bL_{(S,S)}^\top$ and $\bSigma=\bL\bL^\top$ be the Cholesky decomposition of $\bSigma^{(S,S)}$ and $\bSigma$, respectively. It then holds that $\bL_{(S,S)} = \bL^{(S,S)}$ and
\begin{align*}
\bL_{(S,S)}^{-1}\bSigma^{(S,S)}_{\mathbb{E}(\bX|Y)}(\bL_{(S,S)}^{-1})^\top =\{\bL^{-1}\bSigma_{\mathbb{E}(\bX|Y)} (\bL^{-1})^\top\}^{(S,S)} =  \bM^{(S,S)}
\end{align*}
for some ordering of the random vector $\bX$.
\end{lemma}

Then, by similar arguments of \eqref{eq:gep2}, we get
$$\lambda_i(\bSigma^{(S,S)}_{\mathbb{E}(\bX|Y)},\bSigma^{(S,S)})= \lambda_i(\bL_{(S,S)}^{-1}\bSigma^{(S,S)}_{\mathbb{E}(\bX|Y)}(\bL_{(S,S)}^{-1})^\top) = \lambda_i(\bM^{(S,S)})\,,   $$
which gives \eqref{eq:12.1}.

By Assumption (A3), we know there is an event $\Omega$ with probability at least $1-c'p^{-c_2}$ such that on $\Omega$ we have
\begin{align*}
\sup_{S\in\mathcal{S}_k} \max\{\|\widehat\bL^{(S,S)}-\bL^{(S,S)}\|_{\rm op},\|\widehat\bM^{(S,S)}-\bM^{(S,S)}\|_{\rm op}\}\le K\sqrt\frac{k\log p}{n}\,,
\end{align*}
and
\begin{align*}
\sup_{S\in\mathcal{S}_l} \max\{\|\widehat\bL^{(S,S)}-\bL^{(S,S)}\|_{\rm op},\|\widehat\bM^{(S,S)}-\bM^{(S,S)}\|_{\rm op}\}\le K\sqrt\frac{l\log p}{n}\,,
\end{align*}
where $c'>0$ is some constant independent of $S$. On the event $\Omega$ defined above, by Weyl's inequality, we have that
\begin{align*}
\bigg|\sum_{i=1}^d \lambda_i(\widehat\bM^{(S,S)}) - \sum_{i=1}^d \sum_{j\in S\cap \mathcal{A}}\lambda_i (\gamma^{(j,i)})^2 \bigg| &~\overset{\eqref{eq:12.2}}{=} \bigg| \sum_{i=1}^d \{ \lambda_i(\widehat\bM^{(S,S)}) - \lambda_i(\bM^{(S,S)})\}\bigg|\\
&~~~{\le}~Kd\sqrt\frac{k\log p}{n} \overset{\eqref{eq:9.1}}{\le} \frac{d\lambda_d}{16s\mu^2}\,.
\end{align*}
Thus, on the event $\Omega$, if $(S\cap \mathcal{A}) \subsetneq (S'\cap\mathcal{A})$, then
\begin{align}
&~\sum_{i=1}^d \lambda_i(\widehat\bM^{(S,S)})-\sum_{i=1}^d \lambda_i(\widehat\bM^{(S',S')}) \notag\\
\le &~ \sum_{i=1}^d \sum_{j\in S\cap \mathcal{A}}\lambda_i (\gamma^{(j,i)})^2 - \sum_{i=1}^d \sum_{j\in S'\cap \mathcal{A}}\lambda_i (\gamma^{(j,i)})^2 + \frac{d\lambda_d}{8s\mu^2} \notag\\
\le&~ - \sum_{i=1}^d \sum_{j\in (S'\setminus S)\cap \mathcal{A}}\lambda_i (\gamma^{(j,i)})^2 + \frac{d\lambda_d}{8s\mu^2}\,. \label{eq:14.1}
\end{align}
Notice that expression \eqref{eq:8.1} implies that
\begin{align}
\sum_{i=1}^d \lambda_i (\gamma^{(j,i)})^2 \ge \lambda_d \|\bgamma^{(j,\cdot)}\|_2^2 \ge \frac{d\lambda_d}{s\mu^2} \label{eq:14.5.1}
\end{align}
for any $j\in\mathcal{A}$. Then, combining \eqref{eq:14.1} and \eqref{eq:14.5.1}, we have
\begin{align}
\sum_{i=1}^d \lambda_i(\widehat\bM^{(S,S)})<\sum_{i=1}^d \lambda_i(\widehat\bM^{(S',S')})\,.\label{eq:14.6.1}
\end{align}

Given the above expression, we now prove that on the event $\Omega$, for some fixed $j\in\mathcal{A}$ and $j'\notin \mathcal{A}$, it holds that
\begin{align}
q_j\ge q_{j'}\,, \label{eq:15.0}
\end{align}
where $q_k=\mathbb{P}(k\in S_{a,b^*(a)}|\bX)$ for any $k\in[p]$ and some fixed $a\in[A]$. We begin the proof. Define for $\tilde{j}\in\{j,j'\}$ and $b\in[B]$ the following sets:
\begin{align*}
&\mathcal{S}_{b,\tilde{j}}:=\{ (S_{a,1},\ldots,S_{a,B}):b^*(a)=b,\tilde{j}\in S_{a,b} \},\quad\text{and}\\
&~~~~~~~~~\mathcal{S}_{b}:=\{ (S_{a,1},\ldots,S_{a,B}):b^*(a)=b\}\,.
\end{align*}
Let the map $\psi:\mathcal{S}_k\to\mathcal{S}_k$ be defined such that
\begin{align*}
\psi(S):=\bigg\{
\begin{array}{cl}
(S\setminus\{j'\})\cup\{j\} & \text{if}~j'\in S~\text{and}~j\notin S\,,\\
S & \text{otherwise}\,.
\end{array}
\end{align*}
Then for every $S\in\mathcal{S}_k$, either $\psi(S)=S$ or $S\cap\mathcal{A}\subsetneq\psi(S)\cap\mathcal{A}$. Hence, inequality \eqref{eq:14.6.1} implies that
\begin{align}
\sum_{i=1}^d \lambda_i(\widehat\bM^{(S,S)})\le\sum_{i=1}^d \lambda_i(\widehat\bM^{(\psi(S),\psi(S))})\,.\label{eq:15.1.1}
\end{align}
Moreover, by the definition of $\psi(\cdot)$, we know that if $j'\in S$ for some $S\in\mathcal{S}_k$, then $j\in\psi(S)$. Thus, for any $(S_{a,1},\ldots,S_{a,b},\ldots,S_{a,B})\in\mathcal{S}_{b,j'}$, there exists $(S_{a,1},\ldots,\psi(S_{a,b}),\ldots,S_{a,B})\in\mathcal{S}_{b,j}$. To see this clearly, since $(S_{a,1},\ldots,S_{a,b},\ldots,S_{a,B})\in\mathcal{S}_{b,j'}$, then $b^*(a)=b$ and $j'\in S_{a,b}$. Notice that $b^*(a)=b$ implies that
\begin{align*}
&~\max_{b\in[B]}\sum_{i=1}^d \lambda_i(\widehat\bSigma_{\mathbb{E}(\bX|Y)}^{(S_{a,b},S_{a,b})}, \widehat\bSigma^{(S_{a,b},S_{a,b})})\overset{\text{Lemma}~ \ref{lem2}}{=}
\max_{b\in[B]}\sum_{i=1}^d \lambda_i(\widehat\bM^{(S_{a,b},S_{a,b})})\\
\le&~  \sum_{i=1}^d \lambda_i(\widehat\bSigma_{\mathbb{E}(\bX|Y)}^{(S_{a,b},S_{a,b})}, \widehat\bSigma^{(S_{a,b},S_{a,b})})= \sum_{i=1}^d \lambda_i(\widehat\bM^{(S_{a,b},S_{a,b })})\,.
\end{align*}
Together with \eqref{eq:15.1.1}, we know that
\begin{align*}
\max_{b\in[B]}\sum_{i=1}^d \lambda_i(\widehat\bM^{(S_{a,b},S_{a,b})})\le\sum_{i=1}^d \lambda_i(\widehat\bM^{(\psi(S_{a,b}),\psi(S_{a,b }))})\,,
\end{align*}
which implies that $(S_{a,1},\ldots,\psi(S_{a,b}),\ldots,S_{a,B})$ satisfies $b^*(a)=b$. Combining the fact that $j\in\psi(S_{a,b})$, we obtain that $(S_{a,1},\ldots,\psi(S_{a,b}),\ldots,S_{a,B})\in\mathcal{S}_{b,j}$. Hence, $|\mathcal{S}_{b,j'}|\le|\mathcal{S}_{b,j}|$. Then, on $\Omega$, it holds for all $b\in[B]$ that
\begin{align*}
&~\mathbb{P}\{j\in S_{a,b^*(a)} |\bX, b^*(a)=b\}\\
=&~\frac{\mathbb{P}\{j\in S_{a,b^*(a)}, b^*(a)=b|\bX\}}{\mathbb{P}\{b^*(a)=b|\bX\}}=\frac{|\mathcal{S}_{b,j}|}{|\mathcal{S}_{b}|}\\
\ge&~\frac{|\mathcal{S}_{b,j'}|}{|\mathcal{S}_{b}|}=\frac{\mathbb{P}\{j'\in S_{a,b^*(a)}, b^*(a)=b|\bX\}}{\mathbb{P}\{b^*(a)=b|\bX\}}\\
=&~\mathbb{P}\{j'\in S_{a,b^*(a)} |\bX, b^*(a)=b\}\,.
\end{align*}
Therefore, we have proved $q_j\ge q_{j'}$ as shown in \eqref{eq:15.0}.

Notice that
\begin{align*}
\sum_{\tilde{j}\in[p]} q_{\tilde{j}} = \sum_{\tilde{j}\in[p]} \mathbb{P}\{\tilde{j}\in S_{a,b^*(a)}|\bX\}=\frac{1}{B}\sum_{b\in[B]}\sum_{\tilde{j}\in[p]} \mathbb{P}\{\tilde{j}\in S_{a,b^*(a)}|b^*(a)=b,\bX\} = \sum_{\tilde{j}\in[p]} \frac{k}{p} = k\,.
\end{align*}
Thus, by \eqref{eq:15.0} we obtain on $\Omega$ that
\begin{align}
q_j \ge \frac{ \sum_{\tilde{j}\in([p]\setminus\mathcal{A}\cup\{j\})} q_{\tilde{j}} }{p-s+1}=\frac{k-\sum_{\tilde{j}\in\mathcal{A}\setminus\{j\} } q_{\tilde{j}} }{p-s+1} \ge \frac{k-s+1}{p-s+1} \ge \frac{1}{p}\,.  \label{eq:16.0}
\end{align}

Define $\lambda_{a,b;i}:=\lambda_i(\bSigma_{\mathbb{E}(\bX|Y)}^{(S_{a,b},S_{a,b})}, \bSigma^{(S_{a,b},S_{a,b})})$ and $\hat\lambda_{a,b;i}:=\lambda_i(\widehat\bSigma_{\mathbb{E}(\bX|Y)}^{(S_{a,b},S_{a,b})}, \widehat\bSigma^{(S_{a,b},S_{a,b})})$ and $\bv_{a,b;i}:=\bv_i(\bSigma_{\mathbb{E}(\bX|Y)}^{(S_{a,b},S_{a,b})}, \bSigma^{(S_{a,b},S_{a,b})})$ and $\hat\bv_{a,b;i}:=\bv_i(\widehat\bSigma_{\mathbb{E}(\bX|Y)}^{(S_{a,b},S_{a,b})}, \widehat\bSigma^{(S_{a,b},S_{a,b})})$ for $b\in[B]$ and $i\in S_{a,b}$.
Notice that the pair $(\bSigma_{\mathbb{E}(\bX|Y)}^{(S_{a,b},S_{a,b})}, \bSigma^{(S_{a,b},S_{a,b})})$ has $d$ none-zero generalized eigenvalues. Then $\lambda_{a,b;d+1}=\ldots=\lambda_{a,b;k}=0$.
Write $\bV_{a,b} := (\bv_{a,b;1},\ldots,\bv_{a,b:d})$, $\widehat\bV_{a,b} := (\hat\bv_{a,b;1},\ldots,\hat\bv_{a,b:d})$, $\bLambda_{a,b}=\diag(\lambda_{a,b;1},\ldots,\lambda_{a,b;d})$ and $\widehat\bLambda_{a,b}=\diag(\hat\lambda_{a,b;1}-\hat\lambda_{a,b;d+1},\ldots,\hat\lambda_{a,b;d}-\hat\lambda_{a,b;d+1})$. For $\tilde{j} \in S_{a,b^*(a)}$, let
\begin{align*}
\hat{w}_a^{(\tilde{j})} := (\widehat\bV_{a,b^*(a)} \widehat\bLambda_{a,b^*(a)} \widehat\bV_{a,b^*(a)}^{\top})^{(\tilde{j},\tilde{j})} = \sum_{i=1}^d (\hat\lambda_{a,b^*(a);i}-\hat\lambda_{a,b^*(a);d+1})(\hat v_{a,b^*(a);i}^{(\tilde{j})})^2.
\end{align*}
We now give upper bound and lower bound of $\hat{w}_a^{(\tilde{j})}$.

Notice that
\begin{align}
&~\bV_{a,b^*(a)} \bLambda_{a,b^*(a)} \bV_{a,b^*(a)}^{\top}\notag\\ =&~ \sum_{i=1}^d  \lambda_{a,b^*(a);i} \bv_{a,b^*(a);i} \bv_{a,b^*(a);i}^{\top}\overset{(a)}{=} \sum_{i=1}^k  \lambda_{a,b^*(a);i} \bv_{a,b^*(a);i} \bv_{a,b^*(a);i}^{\top} \notag\\ \overset{(b)}{=}&~   \bL_{(S_{a,b^*(a)},S_{a,b^*(a)})}^{-1,\top}\bigg(\sum_{i=1}^k \lambda_{a,b^*(a);i}\bgamma_{a,b^*(a);i}\bgamma_{a,b^*(a);i}^\top\bigg)\bL_{(S_{a,b^*(a)},S_{a,b^*(a)})}^{-1} \notag\\
\overset{(c)}{=}&~\bL_{(S_{a,b^*(a)},S_{a,b^*(a)})}^{-1,\top} (\bL_{(S_{a,b^*(a)},S_{a,b^*(a)})}^{-1}\bSigma_{\mathbb{E}(\bX|Y)}^{(S_{a,b^*(a)},S_{a,b^*(a)})}\bL_{(S_{a,b^*(a)},S_{a,b^*(a)})}^{-1,\top}) \bL_{(S_{a,b^*(a)},S_{a,b^*(a)})}^{-1} \notag\\
\overset{(d)}{=}&~(\bL^{(S_{a,b^*(a)},S_{a,b^*(a)})})^{-1,\top} \bM^{(S_{a,b^*(a)},S_{a,b^*(a)})} (\bL^{(S_{a,b^*(a)},S_{a,b^*(a)})})^{-1} \notag\\
\overset{(e)}{=}&~(\bL^{(S_{a,b^*(a)},S_{a,b^*(a)})})^{-1,\top} (\bgamma^{(S_{a,b^*(a)},\cdot)}\bLambda \bgamma^{(S_{a,b^*(a)},\cdot),\top}) (\bL^{(S_{a,b^*(a)},S_{a,b^*(a)})})^{-1}\,,\label{eq:18.1}
\end{align}
where $\bSigma^{(S_{a,b^*(a)},S_{a,b^*(a)})} = \bL_{(S_{a,b^*(a)},S_{a,b^*(a)})}\bL_{(S_{a,b^*(a)},S_{a,b^*(a)})}^\top$ is the cholesky decomposition of $\bSigma^{(S_{a,b^*(a)},S_{a,b^*(a)})}$ and $\bgamma_{a,b^*(a);i}=\bv_i(\bL_{(S_{a,b^*(a)},S_{a,b^*(a)})}^{-1}\bSigma_{\mathbb{E}(\bX|Y)}^{(S_{a,b^*(a)},S_{a,b^*(a)})}\bL_{(S_{a,b^*(a)},S_{a,b^*(a)})}^{-1,\top})$. Equation $(a)$ is implied by the fact that $\lambda_{a,b^*(a);d+1}=\ldots=\lambda_{a,b^*(a);k}=0$, (b) is obtained by a transformation similar to \eqref{eq:gep2}, (c) is given by the spectral decomposition, (d) uses Lemma \ref{lem2}, and (e) is implied by the fact that $\bM = \bgamma\bLambda\bgamma^\top$.

Let $S_0=S_{a,b^*(a)}\cap\cA$ and $S_0^c=S_{a,b^*(a)}\cap\cA^c$. Without loss of generality, we write
\begin{align*}
\bL^{(S_{a,b^*(a)},S_{a,b^*(a)})}=
\left(\begin{array}{cc}
\bL^{(S_0,S_0)} & \bzero \\
\bL^{(S_0^c,S_0)} & \bL^{(S_0^c,S_0^c)}
\end{array}\right)\,.
\end{align*}
Then, recalling that the transformed basis $\bgamma$ retains the row sparsity of $\bbeta$, we have by equation \eqref{eq:18.1} that
\begin{align*}
\bV_{a,b^*(a)} \bLambda_{a,b^*(a)} \bV_{a,b^*(a)}^{\top}
=&~ \left(\begin{array}{cc}
\bL^{(S_0,S_0),-1,\top} & * \\
\bzero^\top & *
\end{array}\right)
\left(\begin{array}{cc}
\bgamma^{(S_0,\cdot)}\bLambda\bgamma^{(S_0,\cdot),\top} & \bzero \\
\bzero & \bzero
\end{array}\right)
\left(\begin{array}{cc}
\bL^{(S_0,S_0),-1} & \bzero \\
* & *
\end{array}\right) \\
=&~\left(\begin{array}{cc}
\bL^{(S_0,S_0),-1,\top}(\bgamma^{(S_0,\cdot)}\bLambda\bgamma^{(S_0,\cdot),\top}) & \bzero \\
\bzero & \bzero
\end{array}\right)
\left(\begin{array}{cc}
\bL^{(S_0,S_0),-1} & \bzero \\
* & *
\end{array}\right) \\
=&~
\left(\begin{array}{cc}
\bL^{(S_0,S_0),-1,\top}(\bgamma^{(S_0,\cdot)}\bLambda\bgamma^{(S_0,\cdot),\top}) \bL^{(S_0,S_0),-1}& \bzero \\
\bzero & \bzero
\end{array}\right)\,,
\end{align*}
which implies that $(\bV_{a,b^*(a)} \bLambda_{a,b^*(a)} \bV_{a,b^*(a)}^{\top})^{(\tilde{j},\tilde{j})} = (L^{(\tilde{j},\tilde{j})})^{-2}M^{(\tilde{j},\tilde{j})}$ for $\tilde{j}\in S_0$ under some ordering of $\bX$ and $(\bV_{a,b^*(a)} \bLambda_{a,b^*(a)} \bV_{a,b^*(a)}^{\top})^{(\tilde{j},\tilde{j})} = 0 $ otherwise.  Noting that $K\sqrt{k\log(p)/n}\le{\rm min}\{\lambda_1/(4d),\sqrt{\theta_1},\theta_p/(6\sqrt{\theta_1})\}$, on $\Omega$ we have
\begin{align}
&~~| (\widehat\bV_{a,b^*(a)} \widehat\bLambda_{a,b^*(a)} \widehat\bV_{a,b^*(a)}^{\top})^{(\tilde{j},\tilde{j})} - (\bV_{a,b^*(a)} \bLambda_{a,b^*(a)} \bV_{a,b^*(a)}^{\top})^{(\tilde{j},\tilde{j})} | \notag\\
\overset{(1)}{\le} &~ \| \widehat\bV_{a,b^*(a)} \widehat\bLambda_{a,b^*(a)} \widehat\bV_{a,b^*(a)}^{\top} - \bV_{a,b^*(a)} \bLambda_{a,b^*(a)} \bV_{a,b^*(a)}^{\top} \|_{\rm op}
 \notag\\
 =&~ \bigg\| (\widehat{\bL}^{(S_{a,b^*(a)}, S_{a,b^*(a)})})^{-1,\top} \bigg\{\sum_{i=1}^d(\hat\lambda_{a,b^*(a);i} - \hat\lambda_{a,b^*(a);d+1}) \widehat\bgamma_{a,b^*(a);i}\widehat\bgamma_{a,b^*(a);i}^\top \bigg\} (\widehat{\bL}^{(S_{a,b^*(a)}, S_{a,b^*(a)})})^{-1}  - \notag \\
 &~~~(\bL^{(S_{a,b^*(a)},S_{a,b^*(a)})})^{-1,\top}\bigg\{\sum_{i=1}^d (\lambda_{a,b^*(a);i} - \lambda_{a,b^*(a);d+1})\bgamma_{a,b^*(a);i}\bgamma_{a,b^*(a);i}^\top\bigg\}(\bL^{(S_{a,b^*(a)},S_{a,b^*(a)})})^{-1}\bigg\|_{\rm op} \notag\\
\overset{(2)}{\le} &~ 4^{-1} C d \sqrt{\frac{k\log(p)}{n}} \overset{\eqref{eq:9.1}}{\le} \frac{d\tau \lambda_d}{4 s \mu^2}\,, \label{eq:17.1}
  \end{align}
where $\widehat\bgamma_{a,b^*(a);i}=\bv_i((\widehat\bL^{(S_{a,b^*(a)},S_{a,b^*(a)})})^{-1}\widehat\bSigma_{\mathbb{E}(\bX|Y)}^{(S_{a,b^*(a)},S_{a,b^*(a)})}(\widehat\bL^{(S_{a,b^*(a)},S_{a,b^*(a)})})^{-1,\top})$, $\widehat\bSigma = \widehat\bL\widehat\bL^{\top}$ is the Cholesky decomposition of $\widehat\bSigma$ for some ordering of $\bX$. Inequality $(1)$ is obtained by the definition of the
operator norm of a matrix, and (2) is implied by Assumptions (A1), (A1'), (A3), the triangle inequality, and Lemma 2 of \cite{gataric2020sparse}. Thus, on $\Omega \cap \{j \in S_{a,b^*(a)}\}$, we have by Assumption (A4) that
\begin{align}
\hat{w}_a^{(j)} = &~ (\widehat\bV_{a,b^*(a)} \widehat\bLambda_{a,b^*(a)} \widehat\bV_{a,b^*(a)}^{\top})^{({j},{j})} - (\bV_{a,b^*(a)} \bLambda_{a,b^*(a)} \bV_{a,b^*(a)}^{\top})^{({j},{j})} + (\bV_{a,b^*(a)} \bLambda_{a,b^*(a)} \bV_{a,b^*(a)}^{\top})^{({j},{j})} \notag\\
\ge &~ (L^{(j,j)})^{-2}M^{(j,j)} - | (\widehat\bV_{a,b^*(a)} \widehat\bLambda_{a,b^*(a)} \widehat\bV_{a,b^*(a)}^{\top})^{({j},{j})} - (\bV_{a,b^*(a)} \bLambda_{a,b^*(a)} \bV_{a,b^*(a)}^{\top})^{({j},{j})} |  \notag\\
\overset{\eqref{eq:17.1}}{~\ge} & (L^{(j,j)})^{-2}M^{(j,j)} - \frac{d \tau \lambda_d}{4 s \mu^2}\ge \tau \lambda_d \| \bgamma^{(j,\cdot)} \|_2^2 - \frac{d\tau \lambda_d}{4 s \mu^2} \overset{\eqref{eq:8.1}}{\ge} \frac{3 d\tau \lambda_d}{4 s \mu^2}\,. \label{eq:20.1.1}
\end{align}
Similarly, on $\Omega \cap \{j \in S_{a,b^*(a)}\}$, we have \begin{align}\label{eq:20.2.1}
\hat{w}_a^{(j)} \le \frac{5 d \lambda_1 \mu^2}{4 \tau s}\,.
\end{align}
Furthermore, on $\Omega \cap \{ j' \in S_{a,b^*(a)} \}$, it holds that
\begin{align}\label{eq:21.1}
- \frac{d \tau \lambda_d}{4 s \mu^2} \le \hat{w}_a^{(j')} \le \frac{d \tau \lambda_d}{4 s \mu^2}\,.
\end{align}
Finally, for all $j\in[p]$, if $j\notin S_{a,b^*(a)}$, then by the definition of $\hat\bv_{a,b^*(a);i}$ for $i\in[d]$ in Algorithm \ref{alg1}, we know that $(\hat v_{a,b^*(a);i})^{(j)}=0$. Thus, by the definition of $\hat{w}_{a}^{(j)}$, it holds that $\hat{w}_{a}^{(j)}=0$ for $j\notin S_{a,b^*(a)}$.
Hence, we have given upper bound and lower bound of $\hat{w}_a^{(\tilde{j})}$.
Using the lower bound and upper bound given above, we obtain on $\Omega$ that
\begin{align}
&~\mathbb{E}(\hat{w}_a^{(j)}-\hat{w}_a^{(j')} | \bX ) \notag\\
= &~ \mathbb{E}[\hat{w}_a^{(j)}(\mathbf{1}\{j\in S_{a,b^*(a)}\} + \mathbf{1}\{j\notin S_{a,b^*(a)}\} ) | \bX ] - \mathbb{E}[\hat{w}_a^{(j')}(\mathbf{1}\{j'\in S_{a,b^*(a)}\} + \mathbf{1}\{j'\notin S_{a,b^*(a)}\} )  | \bX ] \notag\\
=&~ \mathbb{E}(\hat{w}_a^{(j)}\mathbf{1}\{j\in S_{a,b^*(a)}\}-\hat{w}_a^{(j')}\mathbf{1}\{j'\in S_{a,b^*(a)}\} | \bX )\notag \\
\ge &~ \frac{3 q_j d \tau \lambda_d}{4 s \mu^2} -  \frac{q_{j'}d \tau \lambda_d}{4 s \mu^2} \overset{\eqref{eq:15.0}}{\ge} \frac{ q_j d \tau \lambda_d}{2 s \mu^2} \overset{\eqref{eq:16.0}}{\ge} \frac{ d \tau \lambda_d}{2p s \mu^2}\,. \label{eq:22.1}
\end{align}

Now we let $a$, $j$ and $j'$ freely vary again. Define $\Omega_1 := \{\min_{j\in\mathcal{A}} \hat{w}^{(j)} > \max_{j\in\mathcal{A}^c} \hat{w}^{(j)} \}$.
Let $\bgamma^{(\hat{S},\cdot)} = \bV_d(\bM^{(\hat{S},\hat{S})})$ and $\widehat\bgamma^{(\hat{S},\cdot)}=\bV_d(\widehat\bM^{(\hat{S},\hat{S})})$ denote the $d$ leading eigenvectors of $\bM^{(\hat{S},\hat{S})}$ and $\widehat\bM^{(\hat{S},\hat{S})}$, respectively.
Then, by Assumption (A3) and Theorem 2 of \cite{yu2015useful}, on $\Omega\cap\Omega_1$,
\begin{align*}
L(\widehat\bgamma^{(\hat{S},\cdot)},\bgamma^{(\hat{S},\cdot)})\le \frac{2d^{1/2} \|\widehat\bM^{(\hat{S},\hat{S})}  -\bM^{(\hat{S},\hat{S})} \|_{\rm op} }{\lambda_d} \le  2K \sqrt\frac{dl\log(p)}{n\lambda_d^2}\,.
\end{align*}
Since $l \ge s$, on $\Omega_1$, it holds that $\mathcal{A}\subseteq\hat{S}$, which implies that
$\bgamma = (\bgamma^{(\hat{S},\cdot),\top},\bzero^\top)^\top$. Let $\widehat\bgamma=(\widehat\bgamma^{(\hat{S},\cdot),\top},\bzero^\top)^\top$. We note that the true parameter $\bbeta = (\bgamma^{(\hat{S},\cdot),\top}\bL^{(\hat{S},\hat{S}),-1},\bzero^\top)$ and $\widehat\bbeta = (\widehat\bgamma^{(\hat{S},\cdot),\top}\widehat\bL^{(\hat{S},\hat{S}),-1},\bzero^\top)$ for some ordering of $\bX$. Then, by Assumptions (A1), (A1') and (A3), noting that $K\sqrt{l\log(p)/n}\le{\rm min}\{\lambda_d/(2\sqrt{2}),\sqrt{\theta_1},\theta_p/(6\sqrt{\theta_1})\}$, on $\Omega$ we have
\begin{align*}
L(\bbeta,\widehat\bbeta)= L(\bbeta^{(\hat{S},\cdot)}, \widehat\bbeta^{(\hat{S},\cdot)})=\|\bbeta^{(\hat{S},\cdot)}\bbeta^{(\hat{S},\cdot),\top} - \widehat\bbeta^{(\hat{S},\cdot)}\widehat\bbeta^{(\hat{S},\cdot),\top}\|_{\rm F}\le C \sqrt\frac{dl\log(p)}{n\lambda_d^2}
\end{align*}
for some sufficiently large constant $C>0$.

It suffices to derivative the lower bound of $\mathbb{P}(\Omega\cap\Omega_1)$. Observe that $\hat{w}^{(j)}=A^{-1}\sum_{a=1}^A \hat{w}_a^{(j)}$ for any $j\in[p]$. Then, for any $j\in\mathcal{A}$ and $j'\in\mathcal{A}^c$, on $\Omega$, it holds that
\begin{align*}
&~~~~~~~~\hat{w}^{(j)}-\hat{w}^{(j')} \\
&~~~~= \{ \hat{w}^{(j)} - \mathbb{E}(\hat{w}^{(j)}|\bX) \} - \{ \hat{w}^{(j')} - \mathbb{E}(\hat{w}^{(j')}|\bX) \} +  \mathbb{E}(\hat{w}^{(j)} - \hat{w}^{(j')} |\bX) \\
&~~~\overset{\eqref{eq:22.1}}{\ge}  \frac{ d \tau\lambda_d}{2p s \mu^2} + \{ \hat{w}^{(j)} - \mathbb{E}(\hat{w}^{(j)}|\bX) \} - \{ \hat{w}^{(j')} - \mathbb{E}(\hat{w}^{(j')}|\bX) \}\,,
\end{align*}
which implies
\begin{align*}
\Omega_1^c \subseteq \bigg[\bigcup_{j\in\mathcal{A}} \bigg\{ \hat{w}^{(j)} - \mathbb{E}(\hat{w}^{(j)}|\bX) \le \frac{ d \tau \lambda_d}{4p s \mu^2} \bigg\}\bigg] \cup \bigg[\bigcup_{j'\notin\mathcal{A}} \bigg\{ \hat{w}^{(j')} - \mathbb{E}(\hat{w}^{(j')}|\bX) \ge \frac{ d \tau \lambda_d}{4p s \mu^2} \bigg\}\bigg]\,.
\end{align*}
Notice that conditioned on $\bX$, $\{\hat{w}_a^{(j)}\}_{a\in[A]}$ are i.i.d. random variables, and \eqref{eq:20.1.1}, \eqref{eq:20.2.1} and \eqref{eq:21.1} implies that $\hat{w}_a^{(j)}$ is bounded on $\Omega$ for all $j\in[p]$. Thus, by the union bound and the Hoeffding's inequality conditioned on $\bX$, on $\Omega$ we obtain that
\begin{align*}
\mathbb{P}(\Omega_1^c|\bX) \le p\exp\bigg(-\frac{A\tau^4\lambda_d^2}{50p^2\mu^8\lambda_1^2}\bigg)\,.
\end{align*}
We complete the proof by noting that
\begin{align*}
\mathbb{P}(\Omega\cap\Omega_1) \ge 1 - c' p^{-c_2} - p\exp\bigg(-\frac{A\tau^4\lambda_d^2}{50p^2\mu^8\lambda_1^2}\bigg)\,.
\end{align*}
\hfill$\Box$


\subsection{Proof of Theorem \ref{thm2}}

Consider a specific case where $d=1$ and $H=2$. In this case, the parameter of interest $\bbeta\in\bTheta_{p,d,s}(3)$ reduces to a $1$-dimensional vector $\bbeta_1\in\bTheta_{p,1,s}(3)$. We consider the following structure:
\begin{align}
\bX|(\tilde{Y}=1)\sim\mathcal{N}_p((1-\alpha)\bbeta_1,\bI_p-\bM)\,, ~~~~\mathbb{P}(\tilde{Y}=1)=\alpha\,,\notag\\
\bX|(\tilde{Y}=2)\sim\mathcal{N}_p(-\alpha\bbeta_1,\bI_p-\bM)\,,~~~~ \mathbb{P}(\tilde{Y}=2)=1-\alpha\,,\label{eq:mixgau}
\end{align}
where $\alpha\in(0,1)$ and $\bM$ is a proper positive definite matrix to be defined later.
Under this structure, we have the following three results:
\begin{itemize}
\item[(i)] $\mathbb{E}\bX=\mathbb{E}\{\mathbb{E}(\bX|\tilde{Y})\}=\0$;

\item[(ii)] Define $\bmu_h=\mathbb{E}\{\bX|(\tilde{Y}=h)\}$ and $p_h=\mathbb{P}(\tilde{Y}=h)$ for $h\in[H]$. Then $\mathrm{Cov}\{\mathbb{E}(\bX|\tilde{Y})\}=\sum_{h=1}^H p_h \bmu_h \bmu_h^{\top}-(\sum_{h=1}^H p_h \bmu_h)(\sum_{h=1}^H p_h \bmu_h)^\top=\sum_{h=1}^H p_h \bmu_h \bmu_h^{\top}=\alpha(1-\alpha)\bbeta_1\bbeta_1^{\top}$;

\item[(iii)] Therefore, $\mathrm{Cov}(\bX)=\mathbb{E}(\bX\bX^{\top})=\mathrm{Cov}\{\mathbb{E}(\bX|\tilde{Y})\}+\mathbb{E}\{\mathrm{Cov}(\bX|\tilde{Y})\}=\alpha(1-\alpha)\bbeta_1\bbeta_1^{\top}+\bI_p - \bM$.

\end{itemize}

Letting $\bM = \mathrm{Cov}\{\mathbb{E}(\bX|\tilde{Y})\} = \alpha(1-\alpha)\bbeta_1\bbeta_1^{\top}$, we have $\mathrm{Cov}(\bX)=\bI_p$. Hence, the nonzero generalized eigenvalues of the pair $(\mathrm{Cov}\{\mathbb{E}(\bX|\tilde{Y})\}, \mathrm{Cov}(\bX))$ reduce to the nonzero eigenvalues of $\mathrm{Cov}\{\mathbb{E}(\bX|\tilde{Y})\}$. Recall that $\mathrm{Cov}\{\mathbb{E}(\bX|\tilde{Y})\}=\alpha(1-\alpha)\bbeta_1\bbeta_1^{\top}$, which implies that $\mathrm{Cov}\{\mathbb{E}(\bX|\tilde{Y})\}$ has only one nonzero eigenvalue $\lambda_1=\alpha(1-\alpha)$ and the leading eigenvector, which is the parameter of our interest, corresponds to $\bbeta_1$.

We will use the Fano's lemma \citep{Yu1997} to derive the lower bound. To apply this lemma, we go in three steps: (i) to construct some $\bbeta\in\bTheta_{p,1,s}(3)$; (ii) to derive the Kullback-Leiber divergence between data distributions of interest, \ie, the constructed mixture Gaussian distribution; (iii) to find a subset of $\bTheta_{p,1,s}(3)$ with a proper packing number.

\underline{{\it Proof of Step} (i).} We consider a specific setting of \cite{gataric2020sparse} (Proposition 1) where the parameter of interest is now a vector, not a matrix.
Define $\bu=(s^{-1/2}\mathbf{1}_s^{\top},\mathbf{0}^{\top})^{\top}\in\mathbb{R}^p$, and let $\epsilon\in\big(0,\sqrt{1/(16s)}\big)$. For any $J\in\bTheta_{p-1,1,s-1}:=\{\bV\in\bTheta_{p-1,1}:{\rm supp}(\bV)\le s-1\}$, since $\bu^{\top}\bu=1$, we can select $\widetilde{\bU}\in\bTheta_{p,p}$ such that the first column of $\widetilde{\bU}$ is $\bu$ and the indexes of the nonzero components of $\bbeta_{J}=\widetilde{\bU}(\sqrt{1-\epsilon^2},\epsilon J^{\top})^{\top}$ are a subset of $[s]$. Notice that
\begin{align*}
\bbeta_{J} = \widetilde{\bU}\left(\begin{array}{c}
\sqrt{1-\epsilon^2}\\
\epsilon J
\end{array}\right)=\bu + \widetilde{\bU}\left(\begin{array}{c}
\sqrt{1-\epsilon^2}-1 \\
\epsilon J
\end{array}\right)=:\bu + \widetilde{\bU}\Delta_J\,.
\end{align*}
Due to $\|\widetilde{\bU}\Delta_J\|_{\infty}\le\|\widetilde{\bU}\Delta_J\|_2\le\sqrt{2}\epsilon$ and $\epsilon\le\sqrt{1/(16s)}$, $|\beta_{J}^{(k)}|\in[0.64/\sqrt{s},1.36/\sqrt{s}]$ for any $k\in[s]$, which implies that $\bbeta_J\in\bTheta_{p,1,s}(3)$.

\underline{{\it Proof of Step} (ii).} For any $J, J'\in\bTheta_{p-1,1,s-1}$, let $\bM_J=\lambda_1\bbeta_J\bbeta_J^{\top}$ and $\bM_{J'}=\lambda_1\bbeta_{J'}\bbeta_{J'}^{\top}$, and then denote by $\mathbb{P}(\bbeta_J,\bM_J), \mathbb{P}(\bbeta_{J'},\bM_{J'})$ the corresponding mixture Gaussian distributions specified in \eqref{eq:mixgau}. Write $D_{\mathrm{KL}}(\mathbb{P}\|\mathbb{Q})$ for the KL divergence from a distribution $\mathbb{P}$ to $\mathbb{Q}$. By the convexity of KL divergence, we have
\begin{align}
&~D_{\mathrm{KL}}\big(\mathbb{P}(\bbeta_J,\bM_J)\, \|\, \mathbb{P}(\bbeta_{J'},\bM_{J'})\big) \notag\\
\le &~\alpha D_{\mathrm{KL}}\big(\mathcal{N}_p((1-\alpha)\bbeta_J,\bI_p-\bM_J) \,\|\, \mathcal{N}_p((1-\alpha)\bbeta_{J'},\bI_p-\bM_{J'})\big) \label{eq:dl1}\\
&~+(1-\alpha)D_{\mathrm{KL}}\big( \mathcal{N}_p(-\alpha\bbeta_J,\bI_p-\bM_J)\,\|\, \mathcal{N}_p(-\alpha\bbeta_{J'},\bI_p-\bM_{J'})\big)\,.\label{eq:dl2}
\end{align}
Hence, it suffices to bound the KL divergence between two Gaussian distributions. For the first KL divergence in \eqref{eq:dl1}, we have
\begin{align*}
&~D_{\mathrm{KL}}\big(\mathcal{N}_p((1-\alpha)\bbeta_J,\bI_p-\bM_J) \,\|\, \mathcal{N}_p((1-\alpha)\bbeta_{J'},\bI_p-\bM_{J'})\big)\\
=&~\frac{1}{2}\bigg( \underbrace{[\mathrm{Tr}\{(\bI_p-\bM_{J'})^{-1}(\bI_p-\bM_{J})\} - p]}_{T_1} + \underbrace{\log\bigg\{\frac{\mathrm{det}(\bI_p-\bM_{J'})}{\mathrm{det}(\bI_p-\bM_{J})} \bigg\}}_{T_2} \\
&~ + \underbrace{(1-\alpha)^2(\bbeta_{J'}-\bbeta_J)^{\top}(\bI_p-\bM_{J'})^{-1}(\bbeta_{J'}-\bbeta_J)}_{T_3} \bigg)\,.
\end{align*}
For the first term, it holds that
\begin{align*}
T_1 &~= \mathrm{Tr}\{(\bI_p-\bM_{J'})^{-1}(\bM_{J'}-\bM_J)\}\\
&~= \mathrm{Tr}\bigg\{\bigg(\bI_p + \frac{\lambda_1}{1-\lambda_1}\bbeta_{J'}\bbeta_{J'}^{\top}\bigg)(\lambda_1\bbeta_{J'}\bbeta_{J'}^{\top}-\lambda_1\bbeta_J\bbeta_J^{\top})\bigg\} \\
&~ = \frac{\lambda_1^2}{1-\lambda_1} \{1-\mathrm{Tr}(\bbeta_{J'}\bbeta_{J'}^{\top}\bbeta_J\bbeta_J^{\top})\} \\
&~\le \frac{2\lambda_1^2}{1-\lambda_1}\|\bbeta_J-\bbeta_{J'}\|_2^2\,.
\end{align*}
For the term $T_2$, noting that $\bI_p - \bM_J$ and $\bI_p - \bM_{J'}$ have the same eigenvalues, we then obtain $T_2=0$. For the term $T_3$, we have $T_3 \le (1-\alpha)^2 \lambda_{\max}\{(\bI_p-\bM_{J'})^{-1}\}\|\bbeta_J-\bbeta_{J'}\|_2^2 = (1-\alpha)^2(1-\lambda_1)^{-1}\|\bbeta_J-\bbeta_{J'}\|_2^2$. Combining the upper bounds for the three terms, we finally obtain
\begin{align}\label{eq:dl3}
D_{\mathrm{KL}}\big(\mathcal{N}_p((1-\alpha)\bbeta_J,\bI_p-\bM_J) \,\|\, \mathcal{N}_p((1-\alpha)\bbeta_{J'},\bI_p-\bM_{J'})\big) \le \frac{2\lambda_1^2+(1-\alpha)^2}{2(1-\lambda_1)}\|\bbeta_J-\bbeta_{J'}\|_2^2\,.
\end{align}
The second KL divergence in \eqref{eq:dl2} has a similar upper bound
\begin{align}\label{eq:dl4}
D_{\mathrm{KL}}\big(\mathcal{N}_p(-\alpha\bbeta_J,\bI_p-\bM_J) \,\|\, \mathcal{N}_p(-\alpha\bbeta_{J'},\bI_p-\bM_{J'})\big) \le \frac{2\lambda_1^2+\alpha^2}{2(1-\lambda_1)}\|\bbeta_J-\bbeta_{J'}\|_2^2\,.
\end{align}
Combining \eqref{eq:dl1}-\eqref{eq:dl4}, we have
\begin{align*}
D_{\mathrm{KL}}\big(\mathbb{P}(\bbeta_J,\bM_J)\, \|\, \mathbb{P}(\bbeta_{J'},\bM_{J'})\big)\le \frac{2\lambda_1^2+\lambda_1}{2(1-\lambda_1)}\|\bbeta_J-\bbeta_{J'}\|_2^2
= \frac{2\lambda_1^2+\lambda_1}{1-\lambda_1}(1-\bbeta_J^{\top}\bbeta_{J'})\,.
\end{align*}
Recall $\bbeta_{J}=\widetilde{\bU}(\sqrt{1-\epsilon^2},\epsilon J^{\top})^{\top}$ for any $J\in\bTheta_{p-1,1,s-1}$. Then $\bbeta_J\bbeta_{J'}^{\top} =  1- \epsilon^2 + \epsilon^2 J^{\top}J'\ge 1-\epsilon^2 - \epsilon^2 \|J\|_2\|J'\|_2=1-2\epsilon^2$. Hence,
\begin{align}\label{eq:kl}
D_{\mathrm{KL}}\big(\mathbb{P}(\bbeta_J,\bM_J)\, \|\, \mathbb{P}(\bbeta_{J'},\bM_{J'})\big) \le \frac{2\epsilon^2(2\lambda_1^2+\lambda_1)}{1-\lambda_1}\,.
\end{align}

\underline{{\it Proof of Step} (iii).} Let $\mathcal{J}\subset\bTheta_{p-1,1,s-1}$ such that $\min_{J,J'\in\mathcal{J}} L(\bbeta_J,\bbeta_{J'})\ge c\epsilon$ for some universal constant $c>0$ and $3\le|\mathcal{J}|\le \exp\{n/(4s)\}$, which will be specified later. {Let $\tilde{\bbeta}$ be any possible sparse SIR estimator for $\bbeta$.} Then by \eqref{eq:kl} and Fano's lemma \citep{Yu1997}, we obtain
\begin{align}
\inf_{\tilde{\bbeta}}\sup_{\bbeta\in\bTheta_{p-1,1,s-1}(3)}\mathbb{E}_{P_{\bbeta}}\{L(\tilde{\bbeta},\bbeta)\} \ge&~ \inf_{\tilde{\bbeta}}\max_{J\in\mathcal{J}}\mathbb{E}_{P_{\bbeta_J}}\{L(\tilde{\bbeta},\bbeta_J)\} \notag\\
\ge&~\frac{c\epsilon}{2}\bigg\{ 1 - \frac{{2n\epsilon^2(2\lambda_1^2+\lambda_1)}/{(1-\lambda_1)}+\log 2}{\log |\mathcal{J}|} \bigg\} \notag\\
\ge&~\frac{c\epsilon}{2}\bigg(\frac{1}{3} - \frac{{n\epsilon^2}}{\log |\mathcal{J}|} \bigg) \,,\label{eq:lb}
\end{align}
where we employed the fact that $|\mathcal{J}|\ge 3$ and $\lambda_1=\alpha(1-\alpha)\le 1/4$. Noting that $\log |\mathcal{J}| \le n/(4s)$, we can select $\epsilon^2=\log(|\mathcal{J}|)/(4n)\le 1/(16s)$. Combining \eqref{eq:lb}, we obtain
\begin{align}\label{eq:bd1}
\inf_{\tilde{\bbeta}}\sup_{\bbeta\in\bTheta_{p,1,s}(3)}\mathbb{E}_{P_{\bbeta}}\{L(\tilde{\bbeta},\bbeta)\} \gtrsim \sqrt{\frac{\log|\mathcal{J}|}{n}} \,.
\end{align}

It remains to construct $\mathcal{J}$. Let $\mathcal{S}=\{S\in[p-1]:|S|=s-1\}$. For any $S\in\mathcal{S}$, define $J_S\in\mathbb{R}^{p-1}$ such that $J_S^{(S)}=(s-1)^{-1/2}\mathbf{1}_{s-1}$ and $J_S^{(S^c)}=\mathbf{0}$. Then $J_S\in\bTheta_{p-1,1,s-1}$. By the Varshamov-Gilbert's lemma (Lemma 4.10 in \cite{massart2007concentration}), since $4(s-1)\le p-1$, there exists a subset $\mathcal{T}$ of $\mathcal{S}$ such that
\begin{align}\label{eq:bd2}
\log|\mathcal{T}| =\bigg\lfloor \frac{1}{5}(s-1)\log\bigg( \frac{p-1}{s-1}\bigg) \bigg\rfloor
\end{align}
and $|T\cap T'|\le (s-1)/2$ for any $T,T'\in\mathcal{T}$ with $T\neq T'$. Let $\mathcal{J}=\{J_T:T\in\mathcal{T}\}$, and then $|\mathcal{J}|=|\mathcal{T}|$. Moreover, the fact that $|T\cap T'|\le (s-1)/2$ for any $T,T'\in\mathcal{T}$ with $T\neq T'$ implies that
\begin{align}
\min_{J,J'\in\mathcal{J}:J\neq J'} L(J,J')=\min_{J,J'\in\mathcal{J}:J\neq J'}\sqrt{1-(J^\top J')^2} \ge \sqrt{1 - \bigg(\frac{1}{2}\bigg)^2}\ge\frac{\sqrt{3}}{2}\,.\label{eq:ld}
\end{align}
Notice that
\begin{align*}
L(\bbeta_J,\bbeta_{J'}) = \frac{1}{\sqrt{2}}\|\bbeta_J\bbeta_J^{\top} - \bbeta_{J'}\bbeta_{J'}^{\top}\|_{\mathrm{F}}=\sqrt{\epsilon^4L^2(J,J')+\epsilon^2(1-\epsilon^2)\|J-J'\|_2^2}\ge \epsilon L(J,J')
\end{align*}
for any $J\in\bTheta_{p-1,1,s-1}$. Combining \eqref{eq:ld}, we have
\begin{align*}
\min_{J,J'\in\mathcal{J}:J\neq J'}L(\bbeta_J,\bbeta_{J'}) \ge \frac{\sqrt{3}}{2} \epsilon \,.
\end{align*}
Finally, since $(s-1)\log\{(p-1)/(s-1)\}\ge6$ and $5n\ge4s(s-1)\log\{(p-1)/(s-1)\}$, we have $3\le|\mathcal{J}|\le \exp\{n/(4s)\}$. Therefore, \eqref{eq:bd1} and \eqref{eq:bd2} implies that
\begin{align*}
\inf_{\tilde{\bbeta}}\sup_{\bbeta\in\bTheta_{p,1,s}(3)}\mathbb{E}_{P_{\bbeta}}\{L(\tilde{\bbeta},\bbeta)\} \gtrsim \sqrt{\frac{s\log(p/s)}{n}} \,.
\end{align*}
We complete the proof.\hfill$\Box$

\subsection{Proof of auxiliary lemmas}

\subsubsection{Proof of Lemma \ref{lem1}}
\label{sec:plem1}

Without loss of generality, we assume $\mathbb{E}\bX=\zero$. Then, by definition
\begin{align*}
\bSigma:=&~ \bSigma_{\mathbb{E}(\bX|Y)} = \mathrm{Cov}\{\mathbb{E}(\bX|\tilde{Y})\}=\mathbb{E}[\{\mathbb{E}(\bX|\tilde{Y})\}\{\mathbb{E}(\bX|\tilde{Y})\}^{\top}] \\
=&~\sum_{h=1}^H \mathbb{P}(\tilde{Y}=h)\{\mathbb{E}(\bX|\tilde{Y}=h)\}\{\mathbb{E}(\bX|\tilde{Y}=h)\}^{\top}.
\end{align*}
Notice that
\begin{align*}
\mathbb{E}[\bX\mathbf{1}\{\tilde{Y}=h\}]=&~\mathbb{E}(\mathbb{E}[\bX\mathbf{1}\{\tilde{Y}=h\}|\mathbf{1}\{\tilde{Y}=h\}])\\
=&~\mathbb{E}(\mathbf{1}\{\tilde{Y}=h\}\mathbb{E}[\bX|\mathbf{1}\{\tilde{Y}=h\}])\\
=&~\mathbb{P}[\mathbf{1}\{\tilde{Y}=h\}=1]\cdot\mathbb{E}(\bX|\mathbf{1}\{\tilde{Y}=h\}=1)\\
=&~\mathbb{P}(\tilde{Y}=h)]\cdot\mathbb{E}(\bX|\tilde{Y}=h)\\
=:&~p_h \cdot\mathbb{E}(\bX|\tilde{Y}=h)\,,
\end{align*}
which implies that
\begin{align*}
\bSigma=\sum_{h=1}^H \frac{\mathbb{E}[\bX\mathbf{1}\{\tilde{Y}=h\}](\mathbb{E}[\bX\mathbf{1}\{\tilde{Y}=h\}])^{\top}}{p_h}=:\sum_{h=1}^H \frac{\bu_h\bu_h^{\top}}{p_h}\,.
\end{align*}
By definition, we also have
\begin{align*}
\widehat{\bSigma}:=\widehat{\bSigma}_{\mathbb{E}(\bX|Y)}=\sum_{h=1}^H \frac{\hat\bu_h\hat\bu_h^{\top}}{\hat p_h}\,,
\end{align*}
where $\hat p_h=n^{-1}\sum_{i=1}^n\mathbf{1}\{\tilde{Y}_i=h\}$ and $\hat\bu_h=n^{-1}\sum_{i=1}^n\bX_i\mathbf{1}\{\tilde{Y}_i=h\}$.
Therefore, we obtain
\begin{align*}
\widehat{\bSigma}-\bSigma=&~\sum_{h=1}^H \bigg(\frac{\hat\bu_h\hat\bu_h^{\top}}{\hat p_h}-\frac{\bu_h\bu_h^{\top}}{p_h} \bigg)\\
=&~\sum_{h=1}^H \bigg(\frac{\hat\bu_h\hat\bu_h^{\top}}{\hat p_h}- \frac{\bu_h\bu_h^{\top}}{\hat p_h} + \frac{\bu_h\bu_h^{\top}}{\hat p_h} -\frac{\bu_h\bu_h^{\top}}{p_h} \bigg)\\
=&~\sum_{h=1}^H \{\hat p_h^{-1}(\hat\bu_h\hat\bu_h^{\top}-\bu_h\bu_h^{\top})+(\hat{p}_h^{-1}-p_h^{-1})\bu_h\bu_h^{\top}\}\\
=&~\sum_{h=1}^H \{ p_h^{-1}(\hat\bu_h\hat\bu_h^{\top}-\bu_h\bu_h^{\top})+ (\hat{p}_h^{-1}-p_h^{-1})(\hat\bu_h\hat\bu_h^{\top}-\bu_h\bu_h^{\top}) +(\hat{p}_h^{-1}-p_h^{-1})\bu_h\bu_h^{\top}\}\,.
\end{align*}
For any unit vector $\balpha\in\mathbb{R}^p$, it holds that
\begin{align}
&~\balpha^{\top}(\widehat{\bSigma}-\bSigma)\balpha \notag\\
=&~ \sum_{h=1}^H \{ p_h^{-1} |\balpha^{\top}(\hat\bu_h-\bu_h)|^2 + 2 p_h^{-1} \balpha^{\top}\bu_h(\hat\bu_h-\bu_h)^{\top}\balpha + (\hat{p}_h^{-1}-p_h^{-1})|\balpha^{\top}(\hat\bu_h-\bu_h)|^2 \notag\\
&~ + 2 (\hat{p}_h^{-1}-p_h^{-1}) \balpha^{\top}\bu_h(\hat\bu_h-\bu_h)^{\top}\balpha + (\hat{p}_h^{-1}-p_h^{-1})|\balpha^{\top}\bu_h|^2\}\,.\label{eq:lem1:1}
\end{align}
Now we claim that $\|\bu_h\|_2$ can be bounded by a constant for all $h\in[H]$. By definition,
\begin{align*}
\bSigma = \sum_{h=1}^H \frac{\bu_h\bu_h^{\top}}{p_h} = \bbeta\bLambda\bbeta^{\top},
\end{align*}
where $\bbeta=(\bbeta_1,\ldots,\bbeta_d)^{\top}$ and $\bLambda=\mathrm{diag}\{\lambda_1,\ldots,\lambda_d\}$. Taking trace on both sides, we obtain $\sum_{h=1}^H p_h^{-1}\|\bu_h\|_2^2=\sum_{i=1}^d \lambda_i\le C$ by Assumption (A1), where the constant $C>0$ depends on $d$, $\kappa$ and $\lambda$. Observing that $0<p_h<1$, \eqref{eq:lem1:1} leads to
\begin{align}
&~|\balpha^{\top}(\widehat{\bSigma}-\bSigma)\balpha| \notag\\
\le &~ \sum_{h=1}^H p_h^{-1} |\balpha^{\top}(\hat\bu_h-\bu_h)|^2 + 2 \sum_{h=1}^H p_h^{-1} |(\hat\bu_h-\bu_h)^{\top}\balpha|\cdot |\bu_h|_2 +\sum_{h=1}^H | \hat{p}_h^{-1}-p_h^{-1}|\cdot |\balpha^{\top}(\hat\bu_h-\bu_h)|^2 \notag
\\
&~+ 2 \sum_{h=1}^H | \hat{p}_h^{-1}-p_h^{-1}| \cdot |(\hat\bu_h-\bu_h)^{\top}\balpha| \cdot |\bu_h|_2 + \sum_{h=1}^H  | \hat{p}_h^{-1}-p_h^{-1}| \cdot |\bu_h|_2^2 \notag\\
\lesssim &~ \sum_{h=1}^H  |\balpha^{\top}(\hat\bu_h-\bu_h)|^2 + \sum_{h=1}^H | \hat{p}_h^{-1}-p_h^{-1}|\cdot |\balpha^{\top}(\hat\bu_h-\bu_h)|^2 \notag\\
&~+ \sum_{h=1}^H |(\hat\bu_h-\bu_h)^{\top}\balpha| + \sum_{h=1}^H | \hat{p}_h^{-1}-p_h^{-1}|\cdot |(\hat\bu_h-\bu_h)^{\top}\balpha|+ \sum_{h=1}^H | \hat{p}_h^{-1}-p_h^{-1}|\,. \label{eq:lem1:2}
\end{align}

Consider the term $\balpha^{\top}(\hat\bu_h-\bu_h)$. Let $Z_i(h)=\balpha^{\top}\bX_i\mathbf{1}\{\tilde{Y}_i=h\}-\mathbb{E}[\balpha^{\top}\bX_i\mathbf{1}\{\tilde{Y}_i=h\}]$ for any unit vector $\balpha\in\mathbb{R}^p$ and $h\in[H]$. Then, $\balpha^{\top}(\hat\bu_h-\bu_h)=n^{-1}\sum_{i=1}^n Z_i(h)$.
{Notice that $\bX_i$ is mixture-Gaussian distributed, which implies that $\{Z_i(h)\}_{1=1}^n$ are i.i.d. sub-exponential variables.}
Because $\{\bX_i,\tilde{Y}_i\}$ are i.i.d. random variables and $Z_i(h)$ are functions of $\bX_i$ and $\tilde{Y}_i$, we know that $\{Z_i(h)\}_{1=1}^n$ are also i.i.d. random variables.
By the Bernstein inequality \citep{vershynin2010introduction}, we have
\begin{align}
\mathbb{P}\{|\balpha^{\top}(\hat\bu_h-\bu_h)| \ge t \} \le 2\exp\{-Cn(t^2 \wedge t) \}\,. \label{eq:lem1:3}
\end{align}


For the term $\hat{p}_h^{-1}-p_h^{-1}$, it is easy to check that $\hat{p}_h-p_h=n^{-1}\sum_{i=1}^n \mathbf{1}\{\tilde{Y}_i=h\}-\mathbb{E}[\mathbf{1}\{\tilde{Y}_i=h\}]$, $|\mathbf{1}\{\tilde{Y}_i=h\}-\mathbb{E}[\mathbf{1}\{\tilde{Y}_i=h\}]|\le 1$ and ${\rm Var}(\mathbf{1}\{\tilde{Y}_i=h\}-\mathbb{E}[\mathbf{1}\{\tilde{Y}_i=h\}])\le 1/4$. By the Bernstein inequality \citep{van1996weak}, we have $\mathbb{P}(|\hat{p}_h-p_h|\ge t)\le 2 \exp\{-Cn(t^2 \wedge t) \}$. Let $p_{\min}=\min\{p_1,\ldots,p_H\}$. Due to $|\hat{p}_h^{-1}-p_h^{-1}| = |\hat{p}_h-p_h| /(p_h \hat{p}_h)$, it holds that
\begin{align}
&~\mathbb{P}(|\hat{p}_h^{-1}-p_h^{-1}| \ge 2 p_{\min}^{-2} t) \notag\\
\le&~\mathbb{P}(|\hat{p}_h-p_h|\ge t) + \mathbb{P}(p_h \hat{p}_h \le p_{\min}^2/2) \notag \\
\le &~ \mathbb{P}(|\hat{p}_h-p_h|\ge t) + \mathbb{P}(|\hat{p}_h-p_h| \ge p_{\min}/2) \notag \\
\le&~ 4 \exp\{-Cn(t^2 \wedge t) \} \label{eq:lem1:4}
\end{align}
provided that $p_{\min}\ge 2t$.

Combing \eqref{eq:lem1:2}-\eqref{eq:lem1:4} and using the union bound, we have
\begin{align*}
\mathbb{P}(|\balpha^{\top}(\widehat{\bSigma}-\bSigma)\balpha| \ge t) \lesssim C \exp\{-C'n(t^2 \wedge t^{1/4}) \}.
\end{align*}
By Lemma 2 of \cite{wang2016statistical}, if $\eta>0$ satisfies $k\log(p/\eta)/n\le1$, it then holds that
\begin{align*}
&~ \mathbb{P}\bigg\{ \sup_{\bu\in\mathcal{B}_0^{p-1}(k)} |\balpha^{\top}(\widehat{\bSigma}-\bSigma)\balpha| \ge 2C\sqrt{\frac{k\log(p/\eta)}{n}} \bigg\} \\
\lesssim &~ k^{1/2}\binom{p}{k}\bigg( \frac{128}{\sqrt{255}} \bigg)^{k-1} \mathbb{P}\bigg\{|\balpha^{\top}(\widehat{\bSigma}-\bSigma)\balpha| \ge C\sqrt{\frac{k\log(p/\eta)}{n}}\bigg\} \\
\lesssim &~ k^{1/2}\binom{p}{k}\bigg( \frac{128}{\sqrt{255}} \bigg)^{k-1} \exp\{-C'C^2k\log(p/\eta)\} \\
\lesssim &~ k^{1/2}\binom{p}{k}\bigg( \frac{128}{\sqrt{255}} \bigg)^{k-1} p^{-k} \eta^k ~~~~(\text{choose}~C^2=(C')^{-1})\\
\lesssim &~ \eta\,.
\end{align*}
Similarly, if $k\log(p/\eta)/n>1$, we have
\begin{align*}
\mathbb{P}\bigg[ \sup_{\bu\in\mathcal{B}_0^{p-1}(k)} |\balpha^{\top}(\widehat{\bSigma}-\bSigma)\balpha| \ge 2C \bigg\{\frac{k\log(p/\eta)}{n}\bigg\}^4 \bigg] \lesssim \eta\,.
\end{align*}
Combining the above two equations, we obtain
\begin{align*}
\mathbb{P}\bigg( \sup_{\bu\in\mathcal{B}_0^{p-1}(k)} |\balpha^{\top}(\widehat{\bSigma}-\bSigma)\balpha| \ge 2C \max\bigg[ \bigg\{\frac{k\log(p/\eta)}{n}\bigg\}^4, \sqrt{\frac{k\log(p/\eta)}{n}} \bigg]\bigg) \le C'\eta\,.
\end{align*}
Choose $\eta=p^{-3}$. Noting that $k\log p \ll n$ and \eqref{eq:lem1:4} holds for sufficiently large $n$, we complete the proof. \hfill$\Box$



\subsubsection{Proof of Lemma \ref{lem2}}
\label{sec:plem2}

Without loss of generality, write
\begin{align*}
\bSigma =
\left(\begin{array}{cc}
\bSigma^{(S,S)} & \bSigma^{(S,S^c)} \\
\bSigma^{(S^c,S)} & \bSigma^{(S^c,S^c)}
\end{array}\right)~\text{and}~
\bL =
\left(\begin{array}{cc}
\bL^{(S,S)} & \bzero \\
\bL^{(S^c,S)} & \bL^{(S^c,S^c)}
\end{array}\right)\,.
                        \end{align*}
Then,
\begin{align*}
\left(\begin{array}{cc}
\bSigma^{(S,S)} & \bSigma^{(S,S^c)} \\
\bSigma^{(S^c,S)} & \bSigma^{(S^c,S^c)}
\end{array}\right)=
\left(\begin{array}{cc}
\bL^{(S,S)} & \bzero \\
\bL^{(S^c,S)} & \bL^{(S^c,S^c)}
\end{array}\right)
\left(\begin{array}{cc}
\bL^{(S,S),\top} & \bL^{(S^c,S),\top} \\
 \bzero^\top & \bL^{(S^c,S^c),\top}
\end{array}\right)=
\left(\begin{array}{cc}
\bL^{(S,S)}\bL^{(S,S),\top} & * \\
* & *
\end{array}\right)\,,
\end{align*}
which implies that $\bSigma^{(S,S)} = \bL^{(S,S)}\bL^{(S,S),\top}$. Since $\bSigma^{(S,S)}=\bL_{(S,S)}\bL_{(S,S)}^\top$ and $\bL_{(S,S)}$ is unique, we obtain $\bL_{(S,S)} = \bL^{(S,S)}$.

By the inverse of a block matrix, it holds that
\begin{align*}
\bL^{-1}\bSigma_{\mathbb{E}(\bX|Y)} (\bL^{-1})^\top =&~ \left(\begin{array}{cc}
\bL^{(S,S)} & \bzero \\
\bL^{(S^c,S)} & \bL^{(S^c,S^c)}
\end{array}\right)^{-1}
\left(\begin{array}{cc}
\bSigma_{\mathbb{E}(\bX|Y)}^{(S,S)} & \bSigma_{\mathbb{E}(\bX|Y)}^{(S,S^c)} \\
\bSigma_{\mathbb{E}(\bX|Y)}^{(S^c,S)} & \bSigma_{\mathbb{E}(\bX|Y)}^{(S^c,S^c)}
\end{array}\right)
\left(\begin{array}{cc}
\bL^{(S,S)} & \bzero \\
\bL^{(S^c,S)} & \bL^{(S^c,S^c)}
\end{array}\right)^{-1,\top} \\
=&~\left(\begin{array}{cc}
\bL^{(S,S),-1} & \bzero \\
-\bL^{(S^c,S^c),-1}\bL^{(S^c,S)}\bL^{(S,S),-1} & \bL^{(S^c,S^c),-1}
\end{array}\right)
\left(\begin{array}{cc}
\bSigma_{\mathbb{E}(\bX|Y)}^{(S,S)} & \bSigma_{\mathbb{E}(\bX|Y)}^{(S,S^c)} \\
\bSigma_{\mathbb{E}(\bX|Y)}^{(S^c,S)} & \bSigma_{\mathbb{E}(\bX|Y)}^{(S^c,S^c)}
\end{array}\right)\\
&~\left(\begin{array}{cc}
\bL^{(S,S),-1,\top} & -(\bL^{(S^c,S^c),-1}\bL^{(S^c,S)}\bL^{(S,S),-1})^\top \\ \bzero^\top
 & \bL^{(S^c,S^c),-1,\top}
\end{array}\right) \\
=&~
\left(\begin{array}{cc}
\bL^{(S,S),-1}\bSigma_{\mathbb{E}(\bX|Y)}^{(S,S)} & \bL^{(S,S),-1}\bSigma_{\mathbb{E}(\bX|Y)}^{(S,S^c)} \\
* & *
\end{array}\right)
\left(\begin{array}{cc}
\bL^{(S,S),-1,\top} & * \\ \bzero^\top
 & *
\end{array}\right) \\
=&~\left(\begin{array}{cc}
\bL^{(S,S),-1}\bSigma_{\mathbb{E}(\bX|Y)}^{(S,S)}\bL^{(S,S),-1,\top} & * \\
* & *
\end{array}\right)\,.
\end{align*}
Clearly, $\{\bL^{-1}\bSigma_{\mathbb{E}(\bX|Y)} (\bL^{-1})^\top\}^{(S,S)}=\bL^{(S,S),-1}\bSigma_{\mathbb{E}(\bX|Y)}^{(S,S)}\bL^{(S,S),-1,\top}$. Combining $\bL_{(S,S)} = \bL^{(S,S)}$, we have $\{\bL^{-1}\bSigma_{\mathbb{E}(\bX|Y)} (\bL^{-1})^\top\}^{(S,S)}=\bL_{(S,S)}^{-1}\bSigma^{(S,S)}_{\mathbb{E}(\bX|Y)}(\bL_{(S,S)}^{-1})^\top = \bM^{(S,S)}$. We complete the proof. \hfill$\Box$

%


\singlespacing
\bibliographystyle{JASA}

\bibliography{Bibliography}

\end{document}